\documentclass[11pt,letterpaper]{article}
\usepackage{jheppub}
\usepackage[english]{babel}
\usepackage{amssymb,amsmath}
\usepackage{mathtools}
\usepackage{mathrsfs}
\usepackage{bbold}
\usepackage{hyperref}
\usepackage{empheq}
\usepackage{subfigure}
\usepackage{xfrac} 
\usepackage{graphics,color}
\usepackage{slashed}
\usepackage{graphicx}
\usepackage{braket}
\usepackage{float}
\usepackage{tcolorbox}
\usepackage{booktabs}

\usepackage{tikz}
\usetikzlibrary{arrows,shapes}
\usetikzlibrary{trees}
\usetikzlibrary{matrix,arrows} 				
\usetikzlibrary{positioning}				
\usetikzlibrary{calc,through}				
\usetikzlibrary{decorations.pathreplacing}  
\usepackage{pgffor}							

\usetikzlibrary{decorations.pathmorphing}	
\usetikzlibrary{decorations.markings}
\tikzset{
    vector/.style={decorate, decoration={snake}, draw},
	provector/.style={decorate, decoration={snake,amplitude=2.5pt}, draw},
	antivector/.style={decorate, decoration={snake,amplitude=-2.5pt}, draw},
	graviton/.style={decorate, decoration={snake,amplitude=1.5pt}, draw},
    fermion/.style={draw=black, postaction={decorate},
        decoration={markings,mark=at position .55 with {\arrow[draw=black]{>}}}},
    fermionbar/.style={draw=black, postaction={decorate},
        decoration={markings,mark=at position .55 with {\arrow[draw=black]{<}}}},
    fermionnoarrow/.style={draw=black},
    gluon/.style={decorate, draw=black,
        decoration={coil,amplitude=4pt, segment length=5pt}},
    scalar/.style={dashed,draw=black, postaction={decorate},
        decoration={markings,mark=at position .55 with {\arrow[draw=black]{>}}}},
    scalarbar/.style={dashed,draw=black, postaction={decorate},
        decoration={markings,mark=at position .55 with {\arrow[draw=black]{<}}}},
    scalarnoarrow/.style={dashed,draw=black},
    electron/.style={draw=black, postaction={decorate},
        decoration={markings,mark=at position .55 with {\arrow[draw=black]{>}}}},
	bigvector/.style={decorate, decoration={snake,amplitude=4pt}, draw},
}

\newcommand{\hhrefDoi}[2]{\href{https:/doi.org/#1}{\color{blue}{#2}}}
\newcommand{\hhrefA}[1]{\href{http://arxiv.org/abs/#1}{\color{blue}{arXiv:#1}}}
\newcommand{\hhrefhp}[1]{\href{http://arxiv.org/abs/hep-ph/#1}{\color{blue}{arXiv:hep-ph/#1}}}
\newcommand{\hhrefht}[1]{\href{http://arxiv.org/abs/hep-th/#1}{\color{blue}{arXiv:hep-th/#1}}}

\title{UV Sensitivity of the Axion Mass from Instantons in Partially Broken Gauge Groups}
\preprint{TUM-HEP-1243-19}

\author[a]{Csaba Cs\'aki,}
\author[b]{Maximilian Ruhdorfer,}
\author[c]{and Yuri Shirman}

\emailAdd{csaki@cornell.edu}
\emailAdd{max.ruhdorfer@tum.de}
\emailAdd{yshirman@uci.edu}

\affiliation[a]{Laboratory for Elementary Particle Physics, Cornell University, Ithaca, NY 14853, USA}
\affiliation[b]{Physik-Department, Technische Universit\"{a}t M\"unchen, 85748 Garching, Germany}
\affiliation[c]{Department of Physics and Astronomy, University of California, Irvine, CA 92697, USA}

\abstract{We examine the contribution of small instantons to the axion mass in various UV completions of QCD. We show that the reason behind the potential dominance of such contributions is the non-trivial embedding of QCD into the UV theory. The effects from instantons 
 in the partially broken gauge group appear as ``fractional instanton''  corrections in the effective theory. These will exhibit unusual dependences on the various scales in the problem whenever the index of embedding is non-trivial. We present a full one-instanton calculation of the axion mass in the simplest product group models, carefully keeping track of numerical prefactors. Rather than using a 't Hooft operator approximation we directly evaluate the contributions to the vacuum bubble, automatically capturing the effects of closing up external fermion lines with Higgs loops. This approach is manifestly finite and removes the uncertainty associated with introducing a cutoff scale for the Higgs loops. We verify that the small instantons may dominate over the QCD contribution for very high breaking scales and at least three group factors.}

\makeatletter
\@addtoreset{footnote}{section}
\makeatother

\begin{document}

\maketitle	

\section{Introduction}

The past few decades have seen axions become an ever more important ingredient of modern particle physics beyond the standard model (BSM). The QCD axion provides the most plausible solution to the strong CP problem~\cite{Peccei:1977hh,Peccei:1977ur,Weinberg:1977ma,Wilczek:1977pj}, and at the same time is also a natural dark matter candidate. Besides the QCD axion, axion-like particles (ALPs) are also ubiquitous in string theory, and can be used for many different purposes in BSM model building. For a pedagogic introduction to the axion and the strong CP problem, see for example~\cite{Hook:2018dlk}.

While the coupling of the QCD axion is set by the unknown large Peccei-Quinn (PQ) symmetry breaking scale $f_a$, its mass is surprisingly well predicted. Even though it is due to uncalculable strongly coupled QCD effects, chiral symmetry relates the uncalculable axion mass to the equally uncalculable pion mass, and one obtains the famous relation (see e.g.~\cite{diCortona:2015ldu})
\begin{equation}
m_a^2= \frac{m_u m_d}{(m_u+m_d)^2} \frac{m_\pi^2 f_\pi^2}{f_a^2}
\label{eq:maxion}
\end{equation}
This formula depends only on known IR quantities in addition to the axion decay constant $f_a$ (which sets the coupling strength of the axion), and has been the basis of axion physics ever since the first attempts to directly detect axions. Eq.~(\ref{eq:maxion}) seems like a robust prediction: it is an IR effect where the QCD instantons are strongly coupled, and the expectation is that they will dominate over any additional UV instanton effect, which would be weakly coupled, and hence suppressed. Indeed one can easily check that for the simplest UV completions of QCD the effects of small instantons are strongly suppressed, as long as the theory remains weakly coupled. One possible way to enhance the contributions of small instantons is to change the running of coupling in the UV and make QCD or its UV completion strongly coupled again~\cite{Holdom:1982ex,Holdom:1985vx,Flynn:1987rs}. A particularly elegant realization is to embed QCD into a higher dimensional theory, and it was indeed shown in~\cite{Poppitz:2002ac} that small instanton contributions are naturally enhanced in some 5D theories. This observation allowed~\cite{Gherghetta:2020keg} to construct a 5D model where the axion mass is raised by small instantons.

However, recently Agrawal and Howe (AH) \cite{Agrawal:2017ksf,Agrawal:2017evu} presented the surprising result that for a particular type of UV completion based on product groups small instantons could provide the dominant contribution to the axion mass even if the UV theory remains weakly coupled (and hence fully calculable). This possibility opens up new regions of the parameter space on the  $m_a^2, f_a$ plane. Interesting models implementing the mechanism of~\cite{Agrawal:2017ksf} were proposed in~\cite{Gaillard:2018xgk,Fuentes-Martin:2019dxt,Croon:2019iuh,Reig:2019vqh,Gavela:2018paw},  applications to models of inflation were studied in \cite{Buen-Abad:2019uoc}. 
Other approaches to modify the axion mass within QCD were proposed in \cite{Dine:1986bg,Choi:1988sy,Choi:1998ep,Hook:2018jle} while in \cite{Rubakov:1997vp,Fukuda:2015ana,Berezhiani:2000gh,Hook:2014cda,Blinov:2016kte,Dimopoulos:2016lvn,Gherghetta:2016fhp} the axion mass is raised by coupling it to a new confining gauge group.

In this paper we re-examine the models presented in~\cite{Agrawal:2017ksf,Agrawal:2017evu} in order to identify the underlying dynamics responsible for an  enhancement of small instanton contributions. We identify the non-trivial embedding of QCD into a high-energy gauge group $G$ as the main source behind this enhancement. It is well-known that a spontaneous symmetry breaking can result in unusual  matching relations~\cite{Intriligator:1995id,Csaki:1998vv} of the form 
\begin{equation}
\left( \frac{\Lambda_{G}}{M}\right)^{k b_G} = \left( \frac{\Lambda_{QCD}}{M}\right)^{b_{QCD}} 
\end{equation}
where the integer $k$ is commonly referred to as the index of embedding~\cite{Intriligator:1995id}, $\Lambda_{G}$ and $\Lambda_{QCD}$ ($b_{G}$ and $b_{QCD}$) are the dynamical scales (beta functions) of the high and low energy theories respectively, and $M$ is the symmetry breaking scale. Such a scale matching relation implies that the ordinary 1-instanton solution of the low energy theory is identified with a $k$-instanton solution of the high energy theory~\cite{Csaki:1998vv}. 
 To be more precise, there are certain small instanton configurations that live fully in the broken group, and do not have corresponding instantons in the low energy theory. We will show that the contributions of these configurations to the QCD axion mass scale as 
\begin{equation}
\frac{m_k^2}{m_{QCD}^2} \propto \frac{1}{(4\pi )^F} \left( \frac{\Lambda_{QCD}}{v} \right)^F \left( \frac{M}{\Lambda_{QCD}}\right)^{4-\frac{b_{QCD}}{k}}
\label{eq:finalscaling}
\end{equation}
where $\Lambda_{QCD}$ is the QCD scale, $v$ is the Higgs VEV, $F$ is the number of flavors and $b_{QCD}=\frac{11}{3}N_c -\frac{2}{3} F$. While for $k=1$ every factor in~(\ref{eq:finalscaling}) is smaller than 1 leading to a strong suppression, we find that already for $k=2$ small instanton contributions are enhanced by powers of $M/\Lambda_{QCD}$ that may overwhelm the other suppression factors for sufficiently large $M$.

The aim of this paper is twofold. First we want to explain how~(\ref{eq:finalscaling}) is obtained, and the physics leading to it in terms of the effects of the instantons in the partially broken group. 
Our second aim is to present a detailed estimate of the actual contribution of these instantons to the axion mass. After accounting for all $\mathcal{O}(1)$ factors, including the perturbation of the classical instanton action in the presence of the Higgs VEV\footnote{The importance of this perturbation was also pointed out  by the authors of \cite{Fuentes-Martin:2019dxt} who considered a similar setup.}, we will be able to identify specific models which successfully implement the Agrawal-Howe enhancement mechanism.

The paper is organized as follows: in Sec.~\ref{sec:instantonEstimate} we present a back-of-the-envelope estimate for the scaling of the various small instanton contributions in partially broken gauge theories, and explain why the case with the non-trivial embeddings of the low-energy instanton is the most interesting one. The actual instanton calculation is set up in Sec.~\ref{sec:InstBrokenSUN} where we show how to do the instanton calculation in a completely broken SU(N) theory. Note that in this section we show how to obtain a non-vanishing contribution in the presence of fermions without using the 't Hooft operator approximation, but rather directly performing the integral over the fermionic and bosonic modes, which automatically includes the effects of additional scalar loops closing up the external fermion lines in the 't Hooft operator. We apply these results to the product group theories in Sec.~\ref{sec:ProductGroups} and there we show how much enhancement we can obtain for the axion mass in the various models. We conclude in Sec.~\ref{sec:Conclusions}. We also present two appendices.  App.~\ref{app:tHooft}  contains a detailed description of how to use the 't Hooft operator approximation and a comparison to the full calculation, while in App.~\ref{app:MSbar} we present the conversion from the Pauli-Villars regulator scheme to the commonly used $\overline{\text{MS}}$ scheme.

\section{Small instantons in partially broken groups and index of embedding}
\label{sec:instantonEstimate}

Before diving into the details of the full instanton calculation we would like to present a back-of-the-envelope estimate for the magnitude of the instanton corrections for various UV completions of QCD. There are two novel aspects of the calculation of~\cite{Agrawal:2017ksf} both related to the fact that we are considering small instantons of size $\rho \ll \Lambda_{QCD}^{-1}$.
\begin{itemize}
\item At high energies the Higgs boson(s) become propagating particles allowing us to also consider the effects of closing up the fermion legs of the instanton vertex using Higgs loops (rather than Higgs VEV insertions as is usually done)
\item There may be non-trivial embeddings of QCD into the UV theory where the small instantons of the UV theory correspond to ``fractional instantons'' of QCD.
\end{itemize}
Below we will be estimating the effects of small instantons using both the traditional Higgs VEV insertions as well as the novel loop diagrams. We will see that for the simplest embeddings of  QCD into the UV gauge theory all such effects will be negligible. However we will explain that for the cases with non-trivial embeddings there could be an enhancement by some power of the ratio $M/\Lambda_{QCD}$ which opens up the possibility of these contributions to dominate over the IR contributions of the ordinary QCD instantons. We will show that the examples of small instanton dominance presented in~\cite{Agrawal:2017ksf}  fall in this category of non-trivial embeddings.

Let us assume that the high energy gauge group $G$ is broken to the low energy group $H$ (in phenomenological applications we will, of course, choose $H$ to be $SU(3)_{QCD}$) at the scale $M$ by the VEV of some heavy scalars. 
We will assume that the theory has $F$ flavors of matter fields in the fundamental representation of $G$. In expectation  of our results to the Standard Model we will choose $F$ to be even. In addition, we will introduce gauge singlet scalars $H$ coupled to the matter fields through  Yukawa couplings $y$. These scalars will eventually be identified with the Higgs scalar(s) of the SM. Thus we will assume that in the low energy theory $H$ has both a VEV and  a mass of order $v$. Finally, we will assume that the Yukawa couplings of $H$ are small. This leads us to consider the following hierarchy of scales
\begin{equation}
 yv\ll \Lambda_{QCD}\ll v\ll \Lambda\ll M\,,
\end{equation}
where $\Lambda$ and $\Lambda_{QCD}$ denote RG invariant scales of high and low energy theories respectively. When the embedding of  the low energy group into $G$ is trivial the matching relation between these scales is given by 
\begin{equation}
\label{eq:simplematching}
 \left(\frac{\Lambda_{QCD}}{M}\right)^{b_{QCD}}=\left(\frac{\Lambda}{M}\right)^{b_G}\,.
\end{equation}
Our choice of the hierarchy of scales leads to several important consequences. First, the contributions of the instantons in the broken group (i.e. instantons of size $\rho\lesssim 1/M$) to the effective action are completely calculable. Furthermore, the contributions of small instantons with size $\rho\ll 1/\Lambda_{QCD}$ (and, in particular, of size $\rho\lesssim 1/v$) within the low energy theory but still above the QCD scale are also  calculabe. Finally, the  Higgses $H$ decouple from the low energy physics within the weak coupling regime while the matter fields are effectively massless\footnote{To streamline the analysis we assume here that all the matter fields are  light  compared to $\Lambda_{QCD}$. Accounting for the mass of heavy SM flavors, $t,b,c$, will not affect the relative importance of contributions from different energy scales.} even at the strong coupling scale $\Lambda_{QCD}$.

\begin{figure}
\centering
\subfigure[\label{subfig:tHooftopa}]{\includegraphics[width=0.25\textwidth]{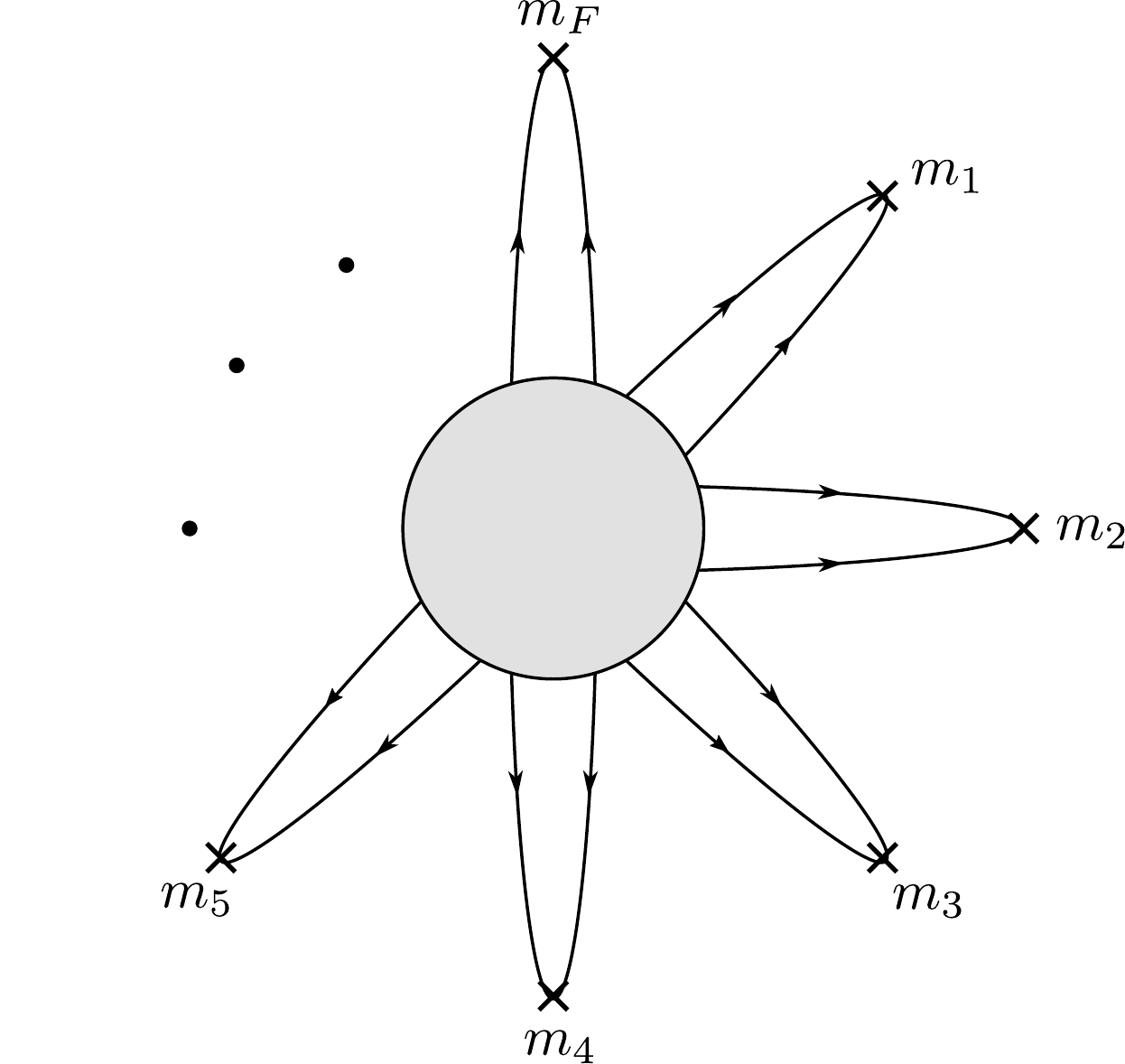}}\hspace*{1cm}
\subfigure[\label{subfig:tHooftopb}]{\includegraphics[width=0.25\textwidth]{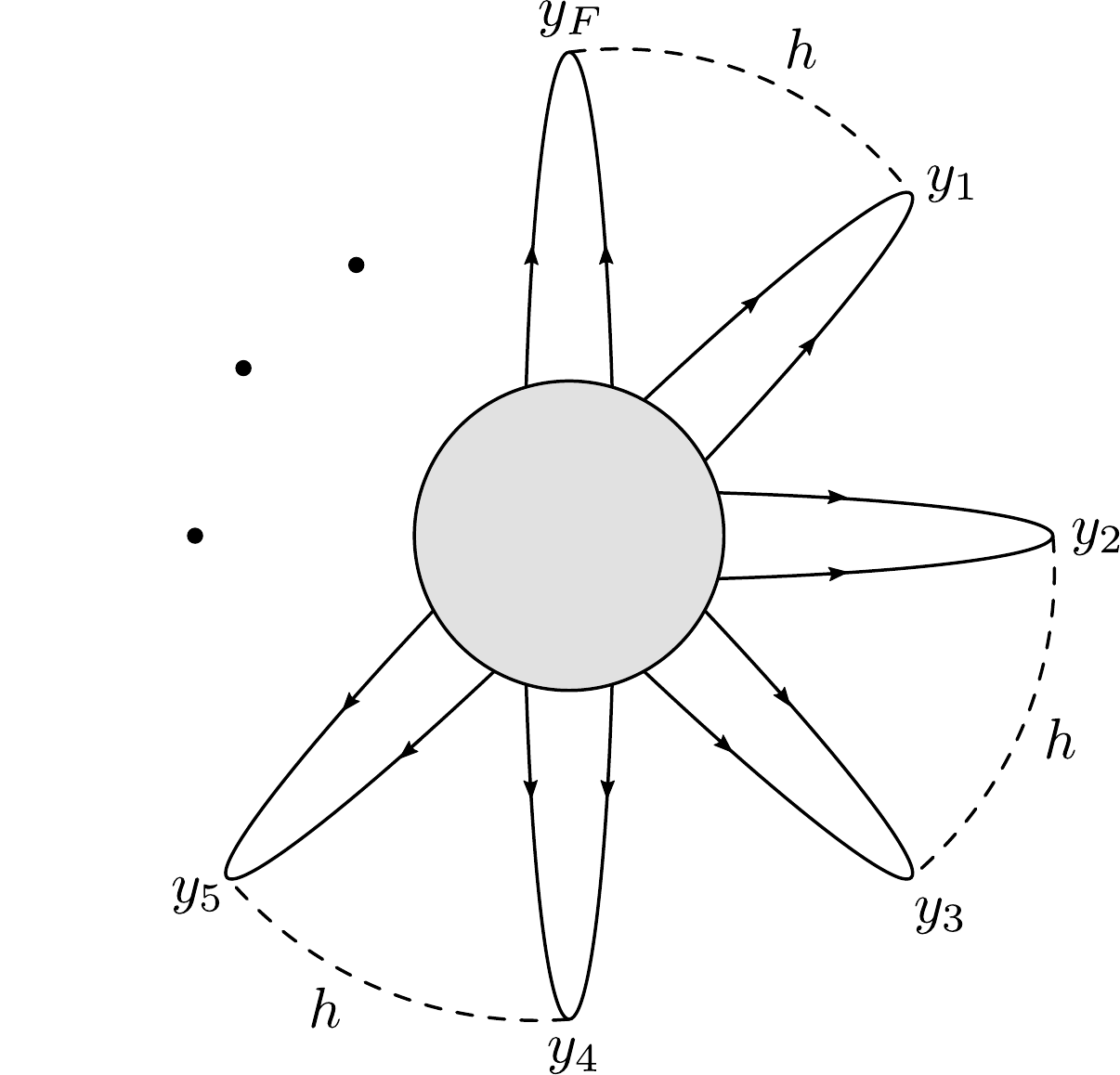}}\hspace*{1cm}
\subfigure[\label{subfig:tHooftopc}]{\includegraphics[width=0.25\textwidth]{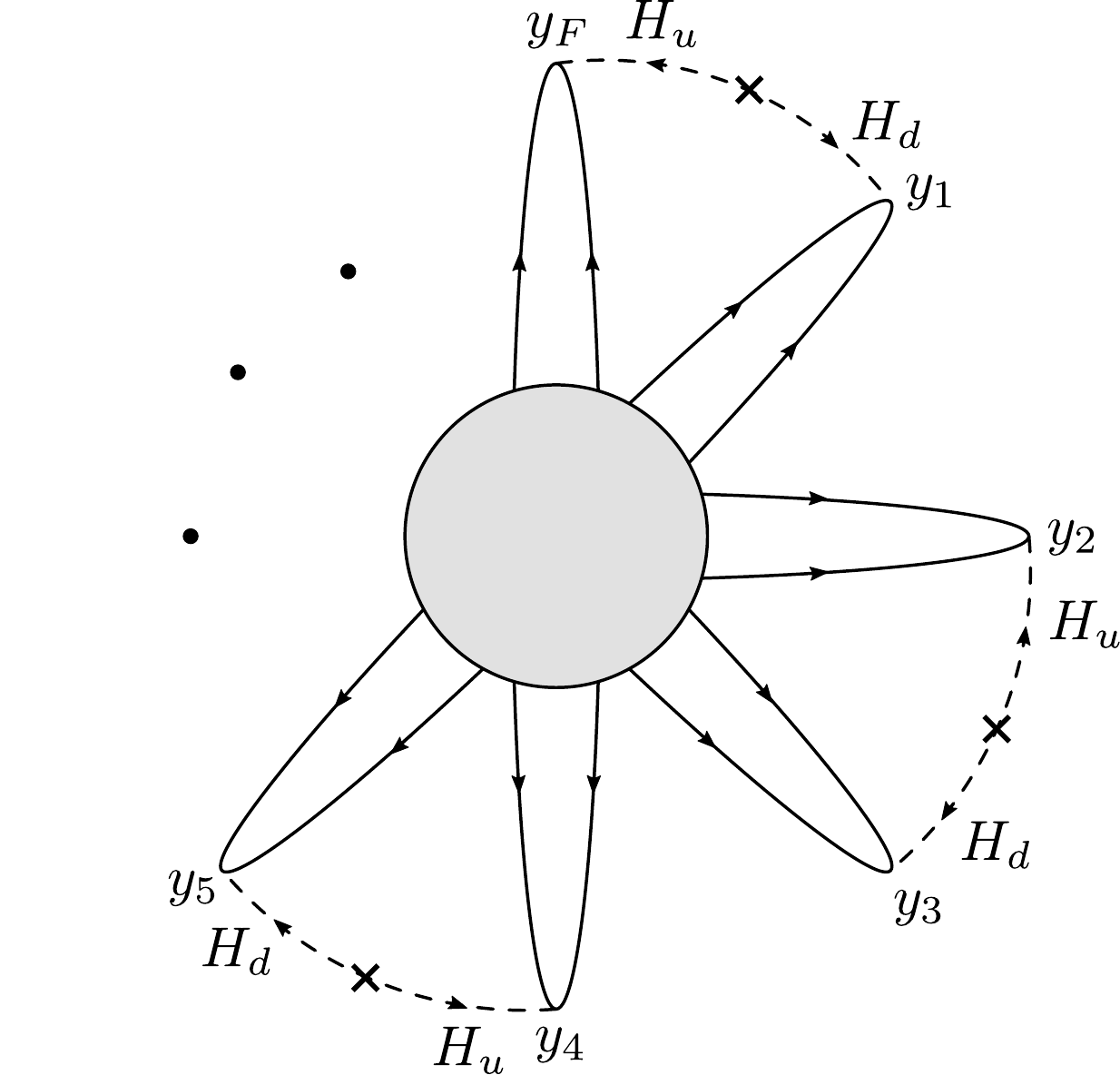}} 
\caption{Contributions to the axion mass obtained from closing up the instanton induced 't Hooft operators. On the left we use Higgs VEV insertions, in the middle we use loops of a dynamical Higgs boson in a single Higgs theory, while on the right we use Higgs loops in a 2HDM. Note that the arrows correspond to chiralities.\label{fig:tHooftop}}
\end{figure}

To obtain a simple estimate of the magnitudes of the effects of the small instantons 
we use an effective Lagrangian below the symmetry breaking scale $M$. Integrating over the instantons of size $\rho<1/M$ generates a 't Hooft operator which must be included in the Lagrangian of the effective theory
\begin{equation}
 \frac{\Lambda^{b_G}}{M^{b_G+3F-4}}\prod_i^F \psi_i\bar\psi_i\,.
\end{equation}
These 't Hooft operators will also contribute to the mass of the axion once the fermion legs are closed up with Higgs VEV insertions or via Higgs loops. 
 Such contributions can be represented by the diagrams in Figure \ref{fig:tHooftop}.
  One important issue to consider is which of these diagrams can possibly contribute to the axion mass. The axion is  the Goldstone boson resulting from the spontaneous breaking of the anomalous $U(1)_{PQ}$ symmetry at a high scale $f_a$. However if the classical action possesses additional exact anomalous unbroken symmetries, one can always redefine the broken $U(1)_{PQ}$ to be anomaly free and the axion remains exactly massless (this is for example the case when one of the SM quarks are exactly massless). As usual the presence of an exact anomalous symmetry will also imply that the QCD $\theta$ angle is unphysical. The Yukawa coupling of the SM fermions breaks any additional global symmetries, hence to obtain a contribution to the axion mass one needs to have a diagram proportional to all SM Yukawa couplings. In models with a single Higgs (like the KSVZ-type  axion models~\cite{Kim:1979if,Shifman:1979if}) 
\begin{equation}
 \sum_{i=1}^{F/2}y H\psi_i\bar\psi_i+\sum_{i=F/2+1}^F y H^\dagger\psi_i\bar\psi_i+\mathrm{h.c.} \label{eq:oneHiggs}
\end{equation}
we can obtain a contribution either through Higgs insertions or via closing up the diagrams using Higgs loops as already depicted 
in Fig.~\ref{subfig:tHooftopa} and \ref{subfig:tHooftopb}.  In other common axion models like DFSZ~\cite{DFSZ,Zhitnitsky:1980tq}  there are two Higgs doublets (2HDM), with Yukawa couplings of the sort
 \begin{equation}
 \sum_{i=1}^{F/2}y H_u\psi_i\bar\psi_i+\sum_{i=F/2+1}^F y H_d\psi_i\bar\psi_i+\mathrm{h.c.} \label{eq:twoHiggs}
\end{equation}
In this case we can still use Higgs insertions, however in order to be able to produce diagrams with Higgs loops one needs an additional $B_\mu$-like term $f_a H_u H_d +h.c.$ Such terms are usually readily present in complete axion models like the DFSZ axion~\cite{DFSZ,Zhitnitsky:1980tq}, and the actual diagram will be of the sort presented in Fig.~\ref{subfig:tHooftopc}.
 The effective theory below $f_a$ will be a one-Higgs doublet model of the sort (\ref{eq:oneHiggs}). As long as $f_a>M$ we can work in the effective one-Higgs doublet model. However if $f_a<M$ one expects the loop diagrams in 2HDMs to be suppressed by powers of $f_a/M$.

We can now compare the contributions to the axion mass from the Higgs VEV insertion diagram 
 \begin{equation}
 \label{eq:smallI-insert}
 m_{M}^2f_a^2=\left(\frac{yv}{M}\right)^F\frac{\Lambda^{b_G}}{M^{b_G-4}}\,,
\end{equation}
to  the contributions from the diagram obtained by closing the Higgs loops:
\begin{equation}
\label{eq:smallI-loop}
 m_{M}^{\prime 2}f_a^2=\left(\frac{y}{4\pi}\right)^F\frac{\Lambda^{b_G}}{M^{b_G-4}}\,.
\end{equation}
where   $m_{M}^2$ and $M_a^{\prime 2}$ represent the contributions of small instantons\footnote{These effects are dominated by instantons of inverse size $M$.} ($\rho\lesssim 1/M$) to the axion mass obtained from VEV-insertion and loop-induced diagrams respectively. For sufficiently large symmetry breaking scale, the suppression of (\ref{eq:smallI-insert}) by $(v/M)^F$ may easily overcome the suppression of the loop-induced contribution by loop factors, making (\ref{eq:smallI-loop}) the dominant contribution from this regime.

There will be similar small instanton contribution to the axion mass from instantons  of size $1/M\lesssim \rho\lesssim 1/v$:
\begin{equation}
 \begin{split}
 \label{eq:smallI-v}
  m_v^2f_a^2&=y^F\frac{\Lambda_{QCD}^{b_{QCD}}}{v^{b_{QCD}-4}}\\
  m_v^{\prime 2}f_a^2&=\left(\frac{y}{4\pi}\right)^F \frac{\Lambda_{QCD}^{b_{QCD}}}{v^{b_{QCD}-4}}\,,
 \end{split}
\end{equation}
where $m_v^2$ and and $m_v^{\prime 2}$ denote the VEV insertion and the Higgs loop induced contributions  respectively\footnote{These effects are dominated by instantons of inverse size $v$.}. It is easy to see that loop-induced contributions in (\ref{eq:smallI-loop})  are small both compared to VEV-insertion and loop-induced contributions in (\ref{eq:smallI-v}).

Below the Higgs mass $v$ the Higgs decouples from the theory and loop-induced contributions are absent. However, given our choice of small Yukawas, the fermions remain light and the instanton diagrams with Higgs VEV insertions still contribute both to the 't Hooft operator and the axion potential. These contributions remain calculable in the $1/v<\rho\ll 1/\Lambda_{QCD}$ regime and at the renormalization scale $\mu$ satisfying $\Lambda_{QCD}\ll\mu\ll v$ are given by
\begin{equation}
\label{eq:axionmu}
 m_\mu^2 f_a^2=\left(y v\right)^F\frac{\Lambda_{QCD}^{b_{QCD}}}{\mu^{b_{QCD}-4}}\,.
\end{equation}
Once again, instanton contributions from lower scales in (\ref{eq:axionmu}) dominate over the instantons of size $1/v$ in (\ref{eq:smallI-v}) and instantons of size $1/M$ in (\ref{eq:smallI-insert})and (\ref{eq:smallI-loop}). As the renormalization scale $\mu$ approaches the actual strong coupling scale $\Lambda_{QCD}$, the perturbative calculation in the one instanton background becomes unreliable. In this regime the contributios of the non-perturbative dynamics to the axion mass are a priori incalculable, however they can be obtained from chiral perturbation theory by relating the axion mass to the pion mass. Nevertheless, one can estimate the final axion mass by taking a naive $\mu\rightarrow \Lambda_{QCD}$ limit:
\begin{equation}
 m_{QCD}^2f_a^2=\left(y v\right)^F\Lambda_{QCD}^{4-F}=m^F\Lambda_{QCD}^{4-F}\,,
\end{equation}
where $F$ is the number of flavors  that remain light at $\Lambda_{QCD}$ and $m_{QCD}$ represents the QCD contribution to the axion mass.

We can now estimate the ratio of loop-induced small instanton and QCD contributions to the axion mass:
\begin{equation}
\frac{M_a^{\prime 2}}{m_a^2} \sim \frac{1}{(4\pi)^F} \left( \frac{\Lambda_{QCD}}{v}\right)^{F}  \left( \frac{\Lambda_{QCD}}{M}\right)^{b_{QCD}-4}\ .
\label{eq:smallinstsimple}
\end{equation}
 As expected the axion mass is dominated by strong coupling QCD contributions while the contributions of small instantons are highly suppressed by powers of $\Lambda/M$ and otherwise are UV independent. Indeed, every term in (\ref{eq:smallinstsimple}) is smaller than one. As a reminder, $b_{QCD}$ is the QCD beta function just below the matching scale with all flavors assumed to be massless: $b_{QCD} = \frac{11}{3} N_c - \frac{2}{3} F = 7$ for QCD with 6 flavors, but most importantly $b_{QCD}>4$ implying a strong suppression by powers of $\Lambda_{QCD}/M$.

There is however an important caveat in the above argument, which is what Agrawal and Howe have exploited \cite{Agrawal:2017evu,Agrawal:2017ksf}. 
The matching relation (\ref{eq:simplematching}) can be modified if the embedding of QCD into the bigger group $G$ is non-trivial. In fact, (\ref{eq:simplematching}) has a very simple and intuitive interpretation: the 1-instanton solution of the low energy  $H$ theory is also a 1-instanton solution of the high energy $G$ theory (with additional bosonic zero modes of the high energy theory lifted by spontaneous symmetry breaking). However, other kinds of embedding are possible~\cite{Csaki:1998vv} -- for example, the 1-instanton solution of the low energy theory may represent a 2-instanton, or in general a $k$-instanton configuration in the high energy theory. The first examples of non-trivial effects due to such instantons were identified in the context of exact results in SUSY gauge theories by Intriligator, Seiberg and Leigh~\cite{Intriligator:1995id}. In this case the matching relation (\ref{eq:simplematching}) would be modified to  
  \begin{equation}
\left( \frac{\Lambda}{M} \right)^{k b_G}= \left( \frac{\Lambda_{QCD}}{M} \right)^{b_{QCD}}
\label{eq:matchingwithk}
\end{equation}
where the integer $k$ is usually referred to as the index of embedding, first identified in ~\cite{Intriligator:1994jr,Intriligator:1994sm,Intriligator:1995id} and explained extensively in~\cite{Csaki:1998vv}. Such a non-trivial factor usually appears when there are instantons in the partially broken gauge group~\cite{Csaki:1998vv}, meaning that the instantons of the unbroken group do not map one-to-one to the instantons of the high energy theory. Topologically it is the homotopy group $\pi_3(G/H)$ that will be relevant, and when both $G$ and $H$ are simple one can show that $\pi_3(G/H)=Z_k$, where $k$ is the index of embedding. In this paper we are interested in models where one breaks a product group to its diagonal subgroup. For example, when the symmetry breaking pattern is given by $SU(N)\times SU(N)\to SU(N)$ the 1-instanton of the low-energy theory actually corresponds to a $(1,1)$ of the UV theory, while for $SU(N)^k \to SU(N)$ the 1-instanton will be a $(1,1,1,\ldots ,1)$ instanton. For the product group case the relevant homotopy group will be $\pi_3(SU(N)^k/SU(N)) = Z\times Z \times \ldots Z$ with $k-1$ $Z$-factors. Either way, if  dynamical scales and beta function coefficients of all UV gauge group factors are equal, the matching relation will be given by Eq.~(\ref{eq:matchingwithk}). More generally the scale matching relation (\ref{eq:matchingwithk}) is replaced by a relation where factors of dynamical scale on the right-hand side are replaced by a product of 1-instanton weights of UV gauge group factors:
\begin{equation}
\label{eq:matchwithkbeta}
 \prod_i^k\left( \frac{\Lambda_i}{M}\right)^{b_i}=\left(\frac{\Lambda_D}{M}\right)^{b_D}\,.
\end{equation}

We can see now how this non-trivial mapping of instantons  (and matching of dynamical scales) would possibly lead to an enhancement of the small instanton contributions. When one has a non-trivial index of embedding, some of the broken instantons are actually topologically distinct from those eventually giving rise to the QCD instanton corrections, hence they will scale differently. From the point of view of scaling they will appear as ``fractional'' $1/k$ instantons, and their contributions may be enhanced compared to the usual QCD instantons. For a case with index of embedding $k$  while the expression of the contribution of the small instantons from the partially broken group are still given by (\ref{eq:smallI-loop}), the use of the modified matching (\ref{eq:matchingwithk}) will result in 
\begin{equation}
\frac{M_a^{\prime 2}}{m_a^2} \sim \frac{1}{(4\pi)^F} \left( \frac{\Lambda_{QCD}}{v}\right)^{F}  \left( \frac{\Lambda_{QCD}}{M}\right)^{\frac{b_{QCD}}{k}-4}\ .
\label{eq:smallinstwithk}
\end{equation}
Already for $k=2$ the sign of the exponent of $\Lambda_{QCD}/M$ will flip, and lead to the possibility of these terms dominating over the ordinary QCD contributions when $M$ is taken to be large. 

In the rest of the paper we will present a detailed calculation of the small instanton effects in the partially broken gauge group to replace (\ref{eq:smallinstwithk}) with a more precise expression, paying careful attention to all the relevant $\mathcal{O}(1)$ factors and perturbation of the classical instanton action in the presence of spontaneous symmetry breaking. This will give us a better understanding of models and parameter regions in which small instanton contributions are dominant.

\section{One instanton effects in a broken \texorpdfstring{$SU(N)$}{SU(N)}}
\label{sec:InstBrokenSUN}
 We now turn to the actual instanton calculation
that will verify the validity of our estimates in Sec.~\ref{sec:instantonEstimate} and provide us with more precise results. As we have explained, contributions of small instantons that are topologically equivalent to single instanton configurations of the low energy theory are always subleading. Instead we will consider instantons of the high energy theory that are absent from the effective low energy description. These instantons must be carefully integrated out and their effects must be taken into account explicitly  when constructing the low energy description. We will be especially interested in product group theories broken to a diagonal subgroup, for example, $SU(N)_1\times SU(N)_2\rightarrow SU(N)_D$.  The low-energy $SU(N)_D$ theory contains only a subset of the instanton solutions of the full $SU(N)_1\times SU(N)_2$ theory \cite{Csaki:1998vv}. For example a 1-instanton configuration in the diagonal subgroup is a $(1,1)$ combination of simultaneous 1-instanton solutions in the individual $SU(N)$ factors. However, configurations with instantons in only one of the $SU(N)$ factors (e.g. $(1,0)$ or $(0,1)$) are absent from the effective theory. Since these instantons are embedded in the completely broken factor of the high energy gauge group, it is useful to review  the instanton calculus in (spontaneously broken) $SU(N)$ gauge theories before re-examining the explicit models in~\cite{Agrawal:2017ksf}. We will loosely follow the instanton calculation in supersymmetric QCD by Cordes~\cite{Cordes:1985um} with slight modifications due to the non-sypersymmetric nature of the problem at hand. It is common practice to perform instanton calculations using Pauli-Villars (PV) regularization, which we will also use here.  However, in perturbative calculations dimensional regularization and the MS or $\overline{\text{MS}}$ scheme are more common. We summarize the formulae needed to convert from PV to $\overline{\text{MS}}$ scheme in App.~\ref{app:MSbar}.

In the following we consider an $SU(N)$ gauge theory with a matter sector consisting of $S \geq N-1$ scalars\footnote{In order to break $SU(N)$ completely one needs at least $N-1$ scalar fields.} $\phi_n, \, n=1,\ldots , S$ and $F$ (approximately) massless fermions $\psi_f, \, f=1,\ldots , F$ in the fundamental representation of $SU(N)$. The euclidean action for this model is given by
\begin{equation}
S_E = S_G + S_\phi + S_\psi \,,
\end{equation}
where
\begin{align}
S_G &= \int d^4 x\, \bigg[\frac{1}{4} G_{\mu\nu}^A G_{\mu\nu}^A + i \theta \frac{g^2}{32\pi^2}  G_{\mu\nu}^A \tilde{G}_{\mu\nu}^A  + \mathcal{L}_{\rm ghost}(\eta,\bar{\eta})\bigg]\,,\\
S_\phi &= \int d^4 x\, \big[ (D_\mu \phi_n)^\dagger D_\mu \phi_n  + V(\phi_n )\big]\,,\\
S_\psi &= \int d^4 x\, \bar{\psi}_f (-i \gamma_\mu D_\mu ) \psi_f\,,
\end{align}
with $D_\mu\phi_n= (\partial_\mu - i g A_\mu^A T^A)\phi_n$. $T^A,\, A=1,\ldots, N^2-1$ are the $SU(N)$ generators. A sum over the scalar and fermion generations is implied. We assume the scalar potential $V(\phi_f)$ to be such that the scalars develop a VEV which breaks the $SU(N)$ gauge symmetry completely.

For vanishing scalar VEVs the euclidean Yang-Mills action $S_G$ possesses exact instanton solutions for the classical equations of motion. The one instanton solution, centered at $x_0$, with unit topological charge ($Q=1$) in singular gauge takes the form \cite{tHooft:1976snw}
\begin{equation}
A_\mu^{Q=1} (x) = \frac{2 \rho^2}{g} \bar{\eta}_{a\mu\nu} \frac{(x-x_0)_v}{(x-x_0)^2 ((x-x_0)^2+\rho^2)} J^a\,,
\label{eq:SUNinstanton}
\end{equation}
where  $\bar{\eta}_{a\mu\nu}$ are 't Hooft symbols, $\rho$ is the instanton size and $J^a,\, a=1,2,3$ are the generators of the $SU(2)$ subgroup into which the instanton is embedded. In the following we will work with the so called minimal embedding, where one embeds the $SU(2)$ into the $2\times 2$ upper-left-hand corner of the $N\times N$ matrices which generate the fundamental representation of $SU(N)$.

Once the scalars obtain a VEV $|\langle \phi_n \rangle | > 0$ and break $SU(N)$ completely, no exact instanton solutions exist\footnote{If $SU(N)$ is only partially broken with an unbroken residual $SU(2)$ subgroup, i.e. rank$(\langle\phi_{i n}\rangle) < N-1$, exact instanton solutions still exist in the unbroken $SU(2)$.}. However, one expects that for small instantons, $g\rho|\langle\phi_n\rangle|\ll 1$, the solution (\ref{eq:SUNinstanton}) remains a good approximation and the path integral is still dominated by instanton-like configurations. The path integral can be performed by using the constrained instanton formalism of Affleck \cite{Affleck:1980mp}. In the constrained instanton formalism the scalars satisfy the equation of motion in the classical instanton background,  $D^2(A_{cl})\phi=0$. As a result to leading order in $g\rho|\langle\phi\rangle|$ the scalar profile is given by
\begin{equation}
\phi_{in} (x)= \begin{cases} \bigg( \frac{x^2}{x^2+\rho^2}\bigg)^{1/2} \langle \phi_{in}\rangle \qquad \text{for }i=1,2\\ \langle \phi_{in}\rangle \,\quad\qquad\qquad\qquad \text{for }i=3,\ldots,N \end{cases}\,,
\label{eq:scalarCl}
\end{equation}
where $i$ is the $SU(N)$ index of the scalar multiplets. 

Evaluating  the classical action of the constrained instanton  with $Q=1$ in the presence of the scalar profile, one finds\footnote{Note that in the background of an instanton with topological charge $Q$, $\tfrac{g^2}{32\pi^2}\int d^4x (G^A_{\mu\nu}\tilde{G}^A_{\mu\nu})_{\rm inst} = Q$.}
\begin{equation}
S_{0} (\rho ) = \frac{8\pi^2}{g^2} + 2\pi^2\rho^2 \sum_{i=1}^2\sum_{n=1}^{S} |\langle \phi_{in} \rangle |^2 + i\theta\,.
\end{equation}
Thus large instantons ($ g\rho  |\langle \phi_f\rangle |\gg 1$) are exponentially suppressed. This provides a natural IR cutoff for the instanton size and makes instanton contributions to observables calculable.

In the following we are interested in $W_{SU(N)}$, the one-instanton semi-classical approximation of the vacuum to vacuum amplitude
\begin{equation}
W_{SU(N)} \equiv \langle 0 | 0 \rangle_{\Delta Q =1} = \frac{\int_{\rm 1-inst} \mathcal{D}A_\mu \mathcal{D}\eta \mathcal{D}\bar{\eta} \mathcal{D}\phi_f \mathcal{D}\phi_f^\dagger \mathcal{D}\psi \mathcal{D}\bar{\psi}_n\, e^{-S_E}}{\int_{A_\mu^{\rm cl}=0} \mathcal{D}A_\mu \mathcal{D}\eta \mathcal{D}\bar{\eta} \mathcal{D}\phi_f \mathcal{D}\phi_f^\dagger \mathcal{D}\psi \mathcal{D}\bar{\psi}_n\, e^{-S_E}}\,.
\label{eq:genFuncAnsatz}
\end{equation}
We can evaluate the functional integral in Eq. (\ref{eq:genFuncAnsatz}) in the semi-classical approximation by expanding the euclidean action to second order in the fields around the classical solutions in Eq. (\ref{eq:scalarCl}) and (\ref{eq:SUNinstanton})
\begin{equation}
S_E = S_0 (\rho ) + \int d^4 x \bigg[ \frac{1}{2} A_\mu M_A A_\mu + \bar{\eta} M_{\rm ghost} \eta + \phi^\dagger M_\phi \phi + \bar{\psi} M_\psi \psi \bigg]\,,
\end{equation}
where $\phi= (\phi_1,\ldots , \phi_{S})^T$ and $\psi = (\psi_1,\ldots , \psi_{F})^T$ are vectors containing all scalar and fermion generations, and perform the functional integral.

The various contributions to the generating functional will be discussed thoroughly in the next sections, but we already present the final expression of the general result here. For the above field content the vacuum to vacuum amplitude is given by
\begin{equation}
\begin{split}
W_{SU(N)} = e^{-i\theta} \int\frac{d^4 x_0 d\rho}{\rho^{5}} d_N(\rho) \int d\tilde{\mu}\, e^{-2\pi^2\rho^2\sum_{i=1}^2\sum_{n=1}^{S} |\langle \tilde{\phi}_{in}\rangle (\tilde{\mu}) |^2}\int \prod_{f=1}^{2F} \rho^{1/2} d\xi_f^{(0)}\,,
\end{split}
\label{eq:SUNFull}
\end{equation}
where $d_N (\rho)$ is the instanton density in vacuum
\begin{equation}
d_N (\rho) = \frac{C_1 e^{- (S - 2 F )\alpha (1/2)}}{(N-1)! (N-2)!}\bigg(\frac{8\pi^2}{g^2}\bigg)^{2N} e^{-\tfrac{8\pi^2}{g^2(1/\rho)} -C_2 N}\,.
\label{eq:instDensity}
\end{equation}
$C_1$ and $C_2$ are defined as
\begin{align}
C_1 &= \frac{2 e^{5/6}}{\pi^2} \approx 0.466\,, \label{eq:C1}\\
C_2 &= \frac{5}{3} \ln 2 - \frac{17}{36} + \frac{1}{3} (\ln 2\pi + \gamma )+ \frac{2}{\pi^2} \sum_{s=1}^{\infty} \frac{\ln s}{s^2} \approx 1.678\,. \label{eq:C2}
\end{align}
$\alpha(t)$ is defined in \cite{tHooft:1976snw} (with $\alpha(0)=0,\,\alpha(1/2)= 0.145873,\,\alpha(1)=0.443307$), $\int d\tilde{\mu}$ is the integral over the collective coordinates corresponding to the orientation of the instanton within $SU(N)$ normalized to unity, and $\int d\xi_f^{(0)}$ is the integral over the fermion zero modes. Note that $\langle\tilde{\phi}_{in}\rangle(\tilde{\mu})$ are the scalar VEVs rotated in group space to account for the arbitrary location of the instanton $SU(2)$ inside $SU(N)$.

\subsection{Bosonic contributions}

Performing the integral in Eq. (\ref{eq:genFuncAnsatz}) over the bosonic sector of the theory, i.e. integrating over the gauge, scalar and ghost fields one obtains 
\begin{equation}
W_{SU(N)} = \int \prod_i d\gamma_i J(\gamma ) e^{-S_0 (\rho )} I_\psi (\gamma ) \frac{(\det' M_A(\gamma ))^{-1/2}  (\det' M_{\rm ghost} (\gamma)) (\det' M_{\phi}(\gamma))^{-2}}{((\det M_A )^{-1/2} (\det M_{\rm ghost})  (\det M_{\phi})^{-2})_{A_\mu^{\rm cl}=0}} \,,
\label{eq:bosonicFI}
\end{equation}
where the contribution from the fermions is encoded in $I_\psi$, which will be computed later, and the determinants $\det'$ are taken over non-zero modes only. 

The zero modes are flat directions in the action and can be parameterised in terms of collective coordinates $\gamma_i$
\begin{equation}
\gamma_i = \begin{cases} (x_0)_i\,\,\,\, i=1,\ldots, 4\\ \rho\qquad \, i=5\\ t^A\quad\,\,\,\, i=A+5 = 6,\ldots , N^2 + 4\end{cases}
\end{equation}
where $x_0$ is the instanton position, $\rho$ its size and $t^A$ are the $N^2-1$ parameters describing general $SU(N)$ transformations. The group theoretic zero modes depend on the embedding of the instanton into $SU(N)$ and their effects can be found by classifying how the generators $T^A$ of the full group transform under the $J^a$ generators of the $SU(2)$ subgroup in which the instanton is embedded. For $SU(N)$ one finds i)  one triplet ($J^a$ themselves) ii) $2(N-2)$ doublets and iii) $N^2-4N +4$ singlets. There are no normalizable zero modes corresponding to singlet generators, which means there are $4N$ normalizable zero modes altogether. 

Replacing the integration over the zero modes in the functional integral by an integration over the collective coordinates introduces the Jacobian $J(\gamma )$ in Eq. (\ref{eq:bosonicFI}). Using the normalization of the zero modes one finds (see e.g. \cite{Cordes:1985um} and \cite{Bernard:1979qt})\footnote{Cordes \cite{Cordes:1985um} and Bernard \cite{Bernard:1979qt} use different normalizations for the $SU(N)$ generators, which is reflected in their different results for the zero-mode normalization (apart from the missing factor of $\rho$ in $||A_\mu^{\rm (isodoub)}||$ in Eq. (5.7) of \cite{Cordes:1985um}, which is clearly a typo). In the following we will follow Cordes' conventions.}
\begin{equation}
\int \prod_i^{4N} d\gamma_i\, J(\gamma ) = \int d^4 x_0\, d\rho\, d\mu\, \frac{2^7}{\rho^5}\bigg( \frac{\pi \rho^2}{g^2}\bigg)^{2N}\,,
\label{eq:zeroModeJacobian}
\end{equation}
where $d\mu$ is the Haar measure of the quotient group $SU(N)/T_N$, with $T_N$ being the stability group of the instanton, i.e. the subgroup of $SU(N)$ that leaves the instanton invariant. In \cite{Cordes:1985um} it is shown that for integrands invariant under $T_N$, the group integration can be expressed as
\begin{equation}
\begin{split}
\int_{SU(N)/T_N} d\mu &= \frac{V(SU(N-1))}{V(T_N)} \int_{SU(N)/SU(N-1)} d\mu \\
&= \frac{2^{4N-6} \pi^{N-2}}{(N-2)!} \int_{S^{2N-1}} \delta( \sqrt{\sum |y_i|^2} -1)\, d^2y_1\ldots d^2 y_N\,.
\end{split}
\label{eq:groupIntMeasure}
\end{equation}
We will denote a general element of the coset $SU(N)/SU(N-1)$ by $\Omega$. It is possible to parameterise $\Omega$ in terms of the $y_i$ \cite{Cordes:1985um}, but the explicit form of $\Omega$ will not be needed in the following. Using the fact that the surface of the $S^{2N-1}$ sphere is given by $S(S^{2N-1}) = 2\pi^N/(N-1)!$, we can define a normalized integration measure
\begin{equation}
\int d\tilde{\mu} = \frac{(N-1)!}{2\pi^N} \int_{S^{2N-1}} \delta( \sqrt{\sum |y_i|^2} -1)\, d^2y_1\ldots d^2 y_N\,.
\label{eq:normIntMeasure}
\end{equation}
As a last step we need to evaluate the functional determinants over the non-zero modes. This calculation has been done in 't Hooft's original paper \cite{tHooft:1976snw} for an $SU(2)$ gauge theory. The generalization to $SU(N)$ is straightforward (see e.g. \cite{Bernard:1979qt}) and yields in Pauli-Villars regularization
\begin{equation}
\begin{split}
&\frac{(\det' M_A(\gamma ))^{-1/2}  (\det' M_{\rm ghost} (\gamma)) (\det' M_{\phi}(\gamma))^{-2}}{((\det M_A )^{-1/2} (\det M_{\rm ghost})  (\det M_{\phi})^{-2})_{A_\mu^{\rm cl}=0}}\\
&= \exp\bigg[ -\big(\tfrac{1}{3}N + \tfrac{1}{6}\sum_t S(t) C(t)\big)\ln(\mu_0\rho) -\alpha (1) -2(N-2)\alpha (1/2) - \sum_t S(t) \alpha (t)\bigg]\,,
\end{split}
\label{eq:bosonicNonZero}
\end{equation}
where $t$ denotes the isospin representation under the instanton $SU(2)$. $S(t)$ is the number of scalar multiplets with isospin $t$, where each complex multiplet counts as 1 and each real multiplet as $1/2$ and $C(t)=\tfrac{2}{3} t (t+1)(2t+1)$. Each scalar fundamental contributes one multiplet in the isospin $1/2$ representation and $(N-2)$ singlets.

Substituting Eqs. (\ref{eq:zeroModeJacobian}), (\ref{eq:groupIntMeasure}), (\ref{eq:normIntMeasure}) and (\ref{eq:bosonicNonZero}) into Eq. (\ref{eq:bosonicFI}) and recalling that in Pauli-Villars regularization each zero-mode yields a factor $\mu_0$ of the regulator field, we obtain
\begin{equation}
\begin{split}
W_{SU(N)} = \frac{C_1\, e^{ - N_S \alpha (1/2)}}{(N-1)!(N-2)!}e^{-i\theta} \bigg(\frac{8\pi^2}{g^2}\bigg)^{2N}  \int \frac{d^4 x_0\, d\rho}{\rho^5} (\mu_0\rho )^{b_0} e^{-8\pi^2/g^2 - C_2 N}\\
\times \int d\tilde{\mu}\, I_\psi (\gamma)\, e^{-2\pi^2\rho^2\sum_{i=1}^2\sum_{n=1}^{S} |\langle \phi_{in}\rangle |^2}\,,
\end{split}
\label{eq:ZSUNBoson}
\end{equation}
where $b_0 = \tfrac{11}{3} N - \tfrac{1}{6}S$ is the bosonic contribution to the $\beta$-function and
\begin{equation}
C_1 = \frac{4\, e^{-\alpha(1) + 4\alpha(1/2)}}{\pi^2}\,,\qquad C_2 = 2\, \ln 2 +2\, \alpha(1/2)\,.
\end{equation}
Note that when using the explicit expression for $\alpha(t)$, the above definition of $C_1$ and $C_2$ agrees with Eqs. (\ref{eq:C1}) and (\ref{eq:C2}).

The group integration $\int d\tilde{\mu}$ in Eq.~(\ref{eq:ZSUNBoson}) corresponds to rotating the instanton embedding in $SU(N)$. This is equivalent to keeping the instanton fixed and instead rotating all other fields, in particular the scalar fields and their VEVs, by a general $SU(N)/SU(N-1)$ group element $\Omega$, i.e. in Eq. (\ref{eq:ZSUNBoson}) we should make the replacement
\begin{equation}
\langle \phi_{in}\rangle \rightarrow \langle \tilde{\phi}_{in}\rangle (\tilde{\mu}) = \sum_{j=1}^N \Omega_{ij} \langle \phi_{jn}\rangle\,.
\label{eq:PhiTilde}
\end{equation}
%

\subsection{Fermionic contributions}

Analogously to the bosonic contributions to the vacuum to vacuum amplitude, one can isolate the zero modes in the integration over the fermionic fields, i.e.
\begin{equation}
\mathcal{D}\psi\mathcal{D}\bar{\psi} = \prod_f ||\psi^{(0)}_f ||^{-1}\, d\xi^{(0)}_f \prod_{f'} ||\bar{\psi}^{(0)}_{f'} ||^{-1}\,  d\bar{\xi}^{(0)}_{f'}\,\mathcal{D}\psi'\, \mathcal{D}\bar{\psi}' \,,
\end{equation}
where $\psi^{(0)}_f$ and $\bar{\psi}_f^{(0)}$ are the zero mode wave functions of the Dirac operator $M^{m n}_\psi = -i \delta^{m n}\gamma_\mu D_\mu$ and $d\xi_f^{(0)}, d\bar{\xi}_f^{(0)}$ are Grassmann integration measures with mass dimension $[d\xi_f^{(0)}]=[ d\bar{\xi}_f^{(0)}] = \tfrac{1}{2}$. The explicit form of the normalized zero-modes in singular gauge, for an instanton centered at $x_0$, is given by \cite{Shifman:2012zz}
\begin{equation}
\psi_f^{(0)}(x)_{\alpha i} = \frac{\rho}{\pi} \frac{(x-x_0)_\mu}{((x-x_0)^2)^{1/2}((x-x_0)^2 + \rho^2)^{3/2}}\begin{pmatrix} 0\\ i (\tau_\mu^+)_i^j \varphi_{\alpha j}\end{pmatrix} \epsilon_{\alpha k}\,,
\label{eq:zeromodesing}
\end{equation}
where $\alpha,i,j=1,2$ are the spinor and $SU(N)$ indices (restricted to the instanton $SU(2)$ with $\psi_f^{(0)}(x)_{\alpha i} =0$ for $i=3,\ldots , N$), respectively.\footnote{Note that the zero modes naively seem to have the wrong dimension (mass dimension $2$ instead of $3/2$), but the combination with the corresponding Grassmann variable $\xi^{(0)}_f$ in the expansion $\psi_f (x) = \sum_{k} \psi^{(k)}_f (x)\, \xi^{(k)}_f$ has the right dimension ($[\xi^{(k)}_f ]=-1/2$, s.t. $\int d\xi^{(k)}_f\, \xi^{(k)}_f = 1$).} $\tau_\mu^+$ is defined as $\tau_\mu^+ = (\vec{\tau},-i)$ with $\vec{\tau}$ being the Pauli matrices. $\varphi_{\alpha j}$ is a two component Weyl spinor with $\varphi_{\alpha j} = \epsilon_{\alpha j}$. Note that for small instantons, far from the instanton center, the zero mode wavefunction is proportional to the free fermion propagator $S_F(x) = \frac{\gamma_\mu x_\mu}{2\pi^2 (x^2)^2}$.
Each massless Dirac fermion in the fundamental representation possesses two zero modes, one for each chirality, in the one instanton background. This implies that in the model with $F$ fermion flavors we have $2F$ fermionic zero modes.

The integral over the non-zero modes can be directly performed, which yields
\begin{equation}
I_\psi = \int \prod_{f=1}^{2F} d\xi_f^{(0)} \mu_0^{-F}\bigg(\frac{\det' M_\psi^\dagger M_\psi}{(\det M_\psi^\dagger M_\psi)_{A_\mu^{\rm cl}=0}}\bigg)^{1/2}\,,
\label{eq:ansatzFermionPI}
\end{equation}
where we assumed normalized zero modes and collectively denoted the zero mode integration measure as $d\xi_f^{(0)}$. Additionally we inserted a factor $\mu_0^{-1/2}$ of the regulator field for each of the $2F$ zero modes, since we work in Pauli-Villars regularization scheme.

The non-zero mode determinant was computed by 't Hooft in his original paper \cite{tHooft:1976snw}
\begin{equation}
\bigg(\frac{\det' M_\psi^\dagger M_\psi}{(\det M_\psi^\dagger M_\psi)_{A_\mu^{\rm cl}=0}}\bigg)^{1/2} = \exp\big[ \tfrac{1}{3} F \ln (\rho \mu_0 ) + 2 F \alpha (1/2)]\,.
\label{eq:nonZeroFermion}
\end{equation}
Combining Eqs. (\ref{eq:ansatzFermionPI}) and (\ref{eq:nonZeroFermion}), we obtain the full fermionic contribution to $W_{SU(N)}$
\begin{equation}
I_\psi = \rho^{F} e^{-\tfrac{2}{3} F \ln (\rho \mu_0 ) + 2 F \alpha (1/2)} \int \prod_{f=1}^{2F} d\xi_f^{(0)}\,.
\label{eq:FermionPI}
\end{equation}
Plugging this result into Eq. (\ref{eq:ZSUNBoson}), one obtains the vacuum to vacuum amplitude for a broken $SU(N)$ gauge theory in a one instanton background, which we already previewed in Eq. ~(\ref{eq:SUNFull}).

\subsection{Vacuum energy/axion potential}
\label{subsec:VacEnergy}
Instanton configurations in the vacuum to vacuum amplitude generate a contribution to the vacuum energy which depends on the $\theta$ angle.  This can be encoded in terms of an effective Lagrangian that captures the 1-(anti)instanton effects in terms of a potential for the $\theta$ angle, which in the presence of an axion will be interpreted as an effective potential/mass term for the axion itself. In a theory without massless fermions this potential is simply given by
\begin{equation}
-\delta\mathcal{L}^{F=0} = 2\int\frac{d\rho}{\rho^5}\int d\tilde{\mu}\, C_N (\rho) \cos (\theta)\,,
\label{eq:thetaPotential}
\end{equation}
where $C_N(\rho )$ contains the instanton density and the action of the Higgs scalars
\begin{equation}
C_N(\rho ) = d_N (\rho)\,  e^{-2\pi^2\rho^2\sum_{i=1}^2\sum_{n=1}^{S} |\langle \tilde{\phi}_{in}\rangle |^2}\,.
\label{eq:CNrho}
\end{equation}
If the theory contains massless fermions, Eq.~(\ref{eq:FermionPI}) implies that due to the $\xi_f^{(0)}$ integration any correlation function, including the vacuum to vacuum amplitude, which does not include the full set of $2F$ chiral fermions vanishes. Effectively the integration projects out the zero mode wave functions, i.e. for a fermion field expanded in eigenmodes $\psi_f = \psi_f^{(0)} \xi_{f}^{(0)} + \ldots$, the integration yields $\int d\xi_f^{(0)} \psi_f = \psi_f^{(0)}$. Thus the effect of massless fermions in the instanton background is captured by an effective $2F$-fermion operator, the so called 't Hooft operator.

However, even in the presence of massless fermions instantons can still generate a potential for the $\theta$ angle if the theory contains further interactions. The easiest way to see that is by working in the effective theory with a 't Hooft operator and closing up the external legs using the additional interaction terms forming a vacuum bubble (see Fig.~\ref{fig:tHooftop}), which contributes to the vacuum energy. Alternatively one can calculate the non-vanishing contribution to the vacuum to vacuum amplitude directly from the path integral by including higher orders in the interaction that includes all massless fermions. In the following we will pursue the second approach, which corresponds to the full calculation. We do expect the effective 't Hooft operator
approach  to be a good approximation to the full calculation, which we will indeed verify in App.~\ref{app:tHooft} where we present the `t Hooft operator method and also compare the results of the two approaches. 

Let us assume the theory contains an additional scalar $H$ (which we will later identify with the SM Higgs), which couples to the massless fermions via Yukawa interactions, i.e. let us add the following term to the Euclidean action
\begin{equation}
\Delta S = S_0 [H] - i \int d^4x \sum_{f=1}^F \frac{y_f}{\sqrt{2}} H(x) \bar{\psi}_f (x) \psi_f (x)\,,
\end{equation}
where $S_0[H]$ is the free action for the scalar $H$. With this addition the vacuum to vacuum amplitude now takes the form
\begin{equation}
\begin{split}
W_{SU(N)} = e^{-i\theta} \int d^4 x_0\int d\tilde{\mu}&\int\frac{d\rho}{\rho^5} C_N(\rho)\int \mathcal{D}H\, e^{-S_0[H]}\\
&\times\int \prod_{f=1}^{F} \rho\, d\xi_f^{(0)} d\bar{\xi}_f^{(0)} e^{i \int d^4 x \sum_{f=1}^{F} \frac{y_f}{\sqrt{2}} H(x) \bar{\psi}_f (x) \psi_f (x)}\,.
\end{split}
\label{eq:MasterEqn}
\end{equation}
At order $F$ in the Yukawa couplings, the expansion of the exponential contains a term with all $2F$ massless fermions. The integration over $\xi_f^{(0)}$ and $\bar{\xi}_f^{(0)}$ projects out the corresponding zero mode wave functions and all lower order terms vanish due to this integration. The leading contribution to $W_{SU(N)}$, assuming $F$ is even so that the path integral of the Higgs field does not vanish (ie. the Higgs loops can be closed up), is therefore\footnote{Note that the $1/F!$ from the expansion of the exponential is compensated by $F!$ terms which are identical after renaming the integration variables.}
\begin{equation}
\begin{split}
W_{SU(N)} = e^{-i\theta} \int d^4 x_0 \int d\tilde{\mu}&\int\frac{d\rho}{\rho^5} C_N(\rho)\int \mathcal{D}H\, e^{-S_0[H]}\prod_{f=1}^{F} \bigg(\frac{i y_f \rho}{\sqrt{2}}\int d^4 x H(x) \bar{\psi}_f^{(0)} (x) \psi_f^{(0)} (x)\bigg)\,.
\end{split}
\end{equation}
Performing the path integral for $H$, only fully contracted Higgs fields survive, each contraction giving a Feynman propagator
\begin{equation}
W_{SU(N)} = e^{-i\theta} \int d^4 x_0 \int d\tilde{\mu}\int\frac{d\rho}{\rho^5} C_N(\rho)\, \kappa_F \prod_{f=1}^{F} \bigg(\frac{y_f \rho}{\sqrt{2}}\bigg) \mathcal{I}^{F/2}\,,
\label{eq:Wferm}
\end{equation}
where $\kappa_F = (F-1)\cdot (F-3)\cdots 1$ counts the number of equivalent contractions and $\mathcal{I}$ is the integral over the fermion zero modes and scalar Feynman propagators $\Delta_F (x_1 - x_2)$
\begin{equation}
\mathcal{I} = -\int d^4 x_1\int d^4x_2\, \bar{\psi}_f^{(0)} (x_1) \psi_f^{(0)} (x_1 ) \bar{\psi}_{f'}^{(0)} (x_2) \psi_{f'}^{(0)} (x_2 ) \Delta_F (x_1 - x_2)\,.
\end{equation}
Using the explicit form for the fermion zero modes\footnote{Note that similarly to the scalars $\phi_{in}$ one should rotate $\psi_f^{(0)}$ with the general $SU(N)/SU(N-1)$ coset element $\Omega$. However, due to the $SU(N)$ invariant Yukawa interaction, the $\Omega$ dependence cancels out and $\mathcal{I}$ is independent of $\tilde{\mu}$.} $\psi_f^{(0)}$ in Eq. (\ref{eq:zeromodesing}) $\mathcal{I}$ simplifies to
\begin{equation}
\mathcal{I} = \frac{\rho^4}{4\pi^8} \int d^4x_1 \int d^4 x_2 \int d^4 p \frac{1}{p^2 +m_H^2} \frac{e^{-ip x_1}}{(x_1^2 + \rho^2)^3} \frac{e^{ip x_2}}{(x_2^2+\rho^2)^3}\,.
\end{equation}
Using the identity
\begin{equation}
\int d^4x \frac{e^{-i p x}}{(x^2+\rho^2)^3} = \frac{\pi^2}{2\rho^2} (p\rho) K_1 (p\rho )\,,
\end{equation}
where $K_1$ is a modified Bessel function of the second kind, we can evaluate $\mathcal{I}$ explicitly in the limit $\rho \ll 1/m_H$
\begin{equation}
\mathcal{I} \simeq \frac{1}{12\pi^2\rho^2}\,.
\end{equation}
Plugging this into Eq. (\ref{eq:Wferm}) we can immediately write down the leading contribution to the potential for the $\theta$ angle, generated by 1-(anti)instanton configurations, for theories with massless fermions and a Yukawa interaction
\begin{equation}
-\delta\mathcal{L}^{F} = 2\int\frac{d\rho}{\rho^5}\int d\tilde{\mu} \, C_N (\rho) \, \kappa_F \prod_{f=1}^{F} \bigg(\frac{y_f}{\sqrt{24}\pi}\bigg) \cos (\theta)\,.
\label{eq:thetaPotentialFerm}
\end{equation}
It is worth emphasizing that $\mathcal{I}$ could be estimated in the effective field theory by soaking up fermion legs of the 't Hooft operator with the Higgs propagators. However, the EFT result would be cutoff dependent while  the above computation is completely convergent and calculable. For more on the correct value of the cutoff scale see App.~\ref{app:tHooft}.

\section{Small instantons in product group models}
\label{sec:ProductGroups}

\begin{figure}[h]
\centering
\includegraphics[width=0.5\textwidth]{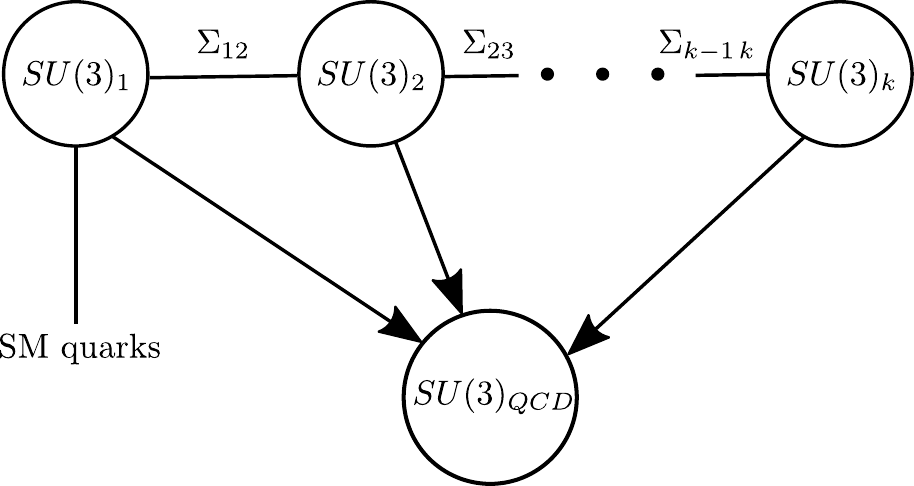}
\caption{Illustration of the product gauge group model introduced in~\cite{Agrawal:2017ksf} to enhance small instanton contributions to the axion mass. The $k$ $SU(3)$ factors are broken at a scale $M$ by scalar link fields $\Sigma_{i\, i+1}$ in the bifundamental representation of $SU(3)_i \times SU(3)_{i+1}$ to their diagonal combination which is identified with $SU(3)_{QCD}$. We further assume that the SM quarks are only charged under $SU(3)_1$.\label{fig:model}}
\end{figure}

Small instanton contributions to the axion mass can dominate over the non-perturbative large QCD instantons in partially broken gauge theories with a non-trivial embedding of $SU(3)_{QCD}$ (see Sec.~\ref{sec:instantonEstimate}). An example of such a setup are the models proposed by Agrawal and Howe \cite{Agrawal:2017ksf,Agrawal:2017evu}, in which a product gauge group consisting of $k $ $SU(3)$ factors is spontaneously broken at a scale $M$ to its diagonal subgroup by $k-1$ link fields $\Sigma_{i\, i+1}$
\begin{equation}
SU(3)_1 \times SU(3)_2 \times \ldots \times SU(3)_k \rightarrow SU(3)_{QCD}\,.
\end{equation}
The diagonal subgroup can then be identified with $SU(3)_{QCD}$. In the following we will assume that all SM quarks are only charged under $SU(3)_1$. For a diagrammatic depiction of the model see Fig.~\ref{fig:model}. 
The individual $SU(3)$ factors by themselves are completely broken and therefore the 1-instanton effects are calculable and finite. The 1-instanton configuration in low energy QCD  corresponds to $k$-instantons of the UV theory with one instanton in each $SU(3)$ factor. In the following we will first discuss some details of the model before we explicitly compute the small instanton contributions to the axion potential in the two simplest realizations with $k=2,3$ and compare the results to~\cite{Agrawal:2017ksf}. Note that in this section we work in Minkowski space.

\subsection{Axions in product group models}

Each of the $SU(3)$ gauge factors comes with its own CP violating $\theta$ angle. Therefore we assume that there is also one anomalous $U(1)_{PQ}$ for each factor, which is spontaneously broken at $f_{a_i} > M$. This yields one axion for each $SU(3)$
\begin{equation}
\mathcal{L} = \sum_{i=1}^k \mathcal{L}_i\,,\qquad \mathcal{L}_i = -\frac{1}{4} G_{i\, \mu\nu}^a G_i^{a\, \mu\nu} + \frac{g_{i}^2}{32\pi^2} \bigg(\frac{a_i}{f_{a_i}} - \theta_i \bigg) G_{i\, \mu\nu}^a \tilde{G}_i^{a\, \mu\nu}\,.
\end{equation}
As depicted in Fig.~\ref{fig:model}, the gauge group is broken to $SU(3)_{QCD}$ by higgsing it with $k-1$ scalar link fields $\Sigma_{i\, i+1}$, which transform as a bifundamental $(3,\bar{3})$ under $SU(3)_i \times SU(3)_{i+1}$. A potential\footnote{One can add $U(1)$ factors to forbid terms like $\mu \det \Sigma$~\cite{Agrawal:2017ksf}.} of the form \cite{Agrawal:2017ksf,Bai:2017zhj} 
\begin{equation}
V(\Sigma ) = -m_\Sigma^2 \text{Tr} (\Sigma \Sigma^\dagger) +\frac{\lambda}{2} [\text{Tr}(\Sigma \Sigma^\dagger )]^2 + \frac{\kappa}{2} \text{Tr}(\Sigma \Sigma^\dagger \Sigma \Sigma^\dagger)
\end{equation}
for each of the link fields induces a VEV
\begin{equation}
\langle\Sigma \rangle = \frac{m_\Sigma}{\sqrt{\kappa + 3 \lambda}} \mathbb{1}_3 \equiv v_\Sigma \mathbb{1}_3\,,
\end{equation}
which for simplicity we take to be the same for all link fields. Each symmetry breaking VEV results in one massive gauge and one massive scalar multiplet in the adjoint representation of the unbroken diagonal group. The masses of gauge and scalar multiplets are of the order\footnote{For simplicity we will assume that $g^2/\kappa\sim 1$ and will not distinguish between the gauge boson and scalar thresholds.} $g_i v_\Sigma$ and $\kappa v_\Sigma$ and they can be integrated out. The dynamical scale of the low energy effective field theory is given by
\begin{equation}
 \Lambda_{QCD}^{b_{QCD}}=\frac{\prod_i^k\Lambda_i^{b_i}}{M^{\sum_i b_i-b_{QCD}}}\,,
\end{equation}
where the matching scale $M$ is the geometric mean of the eigenvalues of the mass matrix for the heavy states. In terms of the QCD coupling constant $g_s$ this implies the usual matching relation at $M$
\begin{equation}
\frac{1}{g_s^2 (M )} = \sum_{i=1}^k \frac{1}{g_i^2(M)}\,,
\label{eq:matching}
\end{equation}
The QCD $\theta$ angle is simply the sum of the individual $SU(3)_i$ $\theta$ angles
\begin{equation}
\bar{\theta}_{QCD} = \sum_{i=1}^k \bar{\theta}_i\,,
\end{equation}
where $\bar{\theta} = \theta + \text{arg det }M_f$ is the physical theta angle, which contains a possible CP violating phase from the fermion mass matrix. At the same time one also has to integrate out the small instantons in the UV theory, which generate a potential for the axions. Thus the effective Lagrangian for the axion fields takes the form\footnote{In~\cite{Agrawal:2017ksf} the mass scale of the potential $m_{a_i}^2 f_{a_i}^2$ was denoted $\Lambda_i^4$.}
 \begin{equation}
\mathcal{L}_{a} = \sum_{i=1}^k m_{a_i}^2 f_{a_i}^2 \cos \bigg(\frac{a_i}{f_{a_i}} - \bar{\theta}_i\bigg) +\frac{g_s^2}{32\pi^2}  \sum_{i=1}^k \bigg(\frac{a_i}{f_{a_i}} - \bar{\theta}_i\bigg)G_{\mu\nu}^a \tilde{G}_i^{\mu\nu} \,.
\label{eq:effLagrangian}
\end{equation}
One can see $\bar\theta_{QCD}$ is relaxed to zero due to two independent effects. First, small instanton contributions in broken gauge factors relax each individual $\bar \theta_i$ to zero. In addition, once QCD confines, the potential is generated for the linear combination $a/f_a=\sum_i a_i/f_{a_i}$ which relaxes $\bar\theta_{QCD}$ to zero just like the usual axion would. In contrast 
 to standard axion models there is not just one but $k$ axions in the IR spectrum and it is the lightest mass eigenstate which plays the role of the QCD axion. When small instanton contributions are dominant the mass of this lightest state can be significantly higher than the standard QCD prediction in Eq.~(\ref{eq:maxion}).
 
\subsection{Small instanton contributions}
When working in the EFT one has to take into account the instanton configurations which are not mapped to the low energy theory, i.e. QCD. These are the independent 1-instanton contributions from $SU(3)_1,\ldots ,SU(3)_k$. Since they are broken to their diagonal combination each $SU(3)$ factor considered separately is completely broken and therefore we can use the formalism of Sec.~\ref{sec:InstBrokenSUN} with three Higgs scalars $\phi_n$, $n=1,2,3$ for each link field, which develop a VEV\footnote{From the point of view of one of the $SU(3)_i$ factors the bifundamental $\Sigma_{i\, i+1}$ looks like three scalars in the fundamental representation.}
\begin{equation}
\langle{\phi_{in}}\rangle = v_\Sigma\, \delta_{in}\,,
\end{equation}
where $i=1,2,3$ are the $SU(3)$ indices. This allows us to evaluate the classical action for the Higgs scalars from one of the link fields in the instanton background explicitly
\begin{equation}
S_0^\phi (\rho) = 2\pi^2\rho^2\sum_{i=1}^2\sum_{n=1}^{3} |\langle \tilde{\phi}_{in}\rangle |^2 = 2\pi^2\rho^2 v_\Sigma^2 \sum_{i=1}^2\sum_{n=1}^{3} |\Omega_{in} |^2 = 4 \pi^2\rho^2 v_\Sigma^2\,,
\label{eq:scalarAction}
\end{equation}
where we considered the rotated VEVs (see Eq. (\ref{eq:PhiTilde})) to account for arbitrary instanton locations inside $SU(3)$.\footnote{Note that the explicit form of $\Omega$ in $SU(3)$ is not needed to obtain the factor of $2$. $\sum_{i=1}^2\sum_{f=1}^{3} |\Omega_{if} |^2$ sums the norms of the first two row vectors in $\Omega$ and since $\Omega\in SU(3)$ each row vector is normalized to unity.} The result is independent of $\tilde{\mu}$ and we can therefore do the now trivial group integration in the results of Section \ref{subsec:VacEnergy}. Note that the scalar action for $SU(3)_2,\ldots , SU(3)_{k-1}$ is twice as large, since each of them couples to two link fields.

We begin by considering the $SU(3)$ sectors without fermions. The last of these sectors, i.e. $SU(3)_k$, has only one scalar link, i.e. $S=3$ scalars in the fundamental representation, and the beta function coefficient $b_k=21/2$. 
For this sector the vacuum-vacuum amplitude contributes directly to the axion potential (see Eq. (\ref{eq:thetaPotential})) with a mass scale $m_{a_k}$ of
\begin{equation}
m_{a_k}^2 f_{a_k}^2 = \left(\frac{\Lambda_k}{M}\right)^{b_k}\left(\frac{M}{2\pi v_\Sigma}\right)^{b_k-4}M^4\,,
\label{eq:Lambdak}
\end{equation}
where the factor $(M/2\pi v_\Sigma)^{b_k-4}$ converts between the physical mass threshold at $M$ and the effective cutoff of the instanton size integral at $1/\rho\sim 2\pi v_\Sigma$, while the RG invariant scale of $SU(3)_k$ sector is defined by
\begin{equation}
 \Lambda_k^{b_k}=\left.d_3(M)\right|_{S_k,F=0} \Gamma [b_k/2 -2] M^{b_k}\,,
\end{equation}
and the instanton weight $d_3(M)|_{S_k,F=0}$ is given in (\ref{eq:instDensity}).
The remaining sectors $i=2,\ldots,k-1$ have two link fields, i.e. $S=6$ scalars in the fundamental representation and the beta function coefficient $b_i=10$. The vacuum-vacuum amplitude contributes to the axion potential in these sectors with a mass scale $m_{a_i}$ which is given by Eq. (\ref{eq:Lambdak}) after the replacement $k\rightarrow i$ and $v_\Sigma\rightarrow \sqrt{2}\, v_\Sigma$. The additional suppression by $2^{2-b_i/2}$ originates from the scalar action which is twice as large, since all of these sectors couple to two link fields.

All the SM quarks are charged under the $SU(3)_1$ sector. Thus its particle content is characterized by $F=6$ approximately massless fermions\footnote{To a good approximation all SM quarks are massless at scales $M \gg $ TeV.} and $S = 3$ scalars in the fundamental representation, corresponding to a beta function coefficient of $b_1 = 13/2$. Taking the result for the vacuum energy in the instanton background with massless quarks and a Yukawa interaction from Eq. (\ref{eq:thetaPotentialFerm}) for $N=3$ and $\theta = \bar{\theta}_1 - \tfrac{a_1}{f_{a_1}}$ and matching it to the axion potential in Eq. (\ref{eq:effLagrangian}) we obtain the scale $m_{a_1}^2 f_{a_1}^2$
\begin{equation}
m_{a_1}^2 f_{a_1}^2 = K \int\frac{d\rho}{\rho^5} \, 2 C_3 (\rho) \,.
\label{eq:Lambda1intermediate}
\end{equation}
where $K$ is given by
\begin{equation}
K = \frac{40}{9}\frac{y_u y_d y_s y_c y_b y_t}{(16\pi^2)^3}\,.
\end{equation}
Note that $K$ reproduces a loop factor expected from an EFT diagram in 
Fig.~\ref{subfig:tHooftopb} 
and included in the results of~\cite{Agrawal:2017ksf}. However, the full calculation of correlation functions in the instanton background performed in Sec.~\ref{subsec:VacEnergy} allows us to extract the exact numerical coefficient multiplying this loop factor. Performing the $\rho$ integral in Eq. (\ref{eq:Lambda1intermediate}) we find
\begin{equation}
m_{a_1}^2 f_{a_1}^2 = K \left(\frac{\Lambda_1}{M}\right)^{b_1}\left(\frac{M}{2\pi v_\Sigma}\right)^{b_1-4} M^4\,,
\label{eq:Lambda1}
\end{equation}
where the dynamical scale of $SU(3)_1$ is defined by
\begin{equation}
 \Lambda_1^{b_1}=\left.d_3(M)\right|_{S=3,F=6} \Gamma [b_1/2 -2] M^{b_1}\,,
\label{eq:Lambda1}
\end{equation}
and once again the instanton weight $d_3(M)|_{S=3,F=6}$ is given in (\ref{eq:instDensity}).
Note that these results are in agreement with the qualitative discussion of Sec.~\ref{sec:instantonEstimate}.  

The unusual scaling of the axion mass with the physical QCD scale can be seen from the fact that $d_3(M) \sim \exp(-\frac{8\pi^2}{g_i^2(M)})$ where $g_i^2$ is the coupling of the i$^{th}$ SU(3) factor rather than the actual QCD coupling, implying that $\Lambda_i^{b_i}$ will be a fractional power of $\Lambda_{QCD}^{b_{QCD}}$, where the actual fraction depends on the ratios of coupling strengths and the distribution of the matter fields among the different group factors.

However the full expression of the corrections to the axion mass Eqs.~(\ref{eq:Lambdak})-(\ref{eq:Lambda1}) also includes an additional suppression factors, for example the conversion factor $(M/2\pi v_\Sigma)^{b_i-4}$. Indeed the presence of this factor implies that, up to an order one coefficient, our results for $m_{a_i}^2f_{a_i}^2$ are smaller than the previous estimates ($\tilde m_{a_i}^2f_{a_i}^2$) in \cite{Agrawal:2017ksf} by a factor of 
\begin{equation}
\frac{m_{a_i}^2}{\tilde m_{a_i}^2}\simeq 2^{-6} \cdot \bigg(\frac{M}{2\pi v_\Sigma}\bigg)^{b_i -4}\,.
\label{eq:suppressionFactor}
\end{equation}
This suppression is due to two independent reasons:
\begin{itemize}
\item Our vacuum instanton density $d_N (\rho)|_{S=F=0}$ is smaller by a factor of $2^{-2N}$ than the one used in \cite{Agrawal:2017ksf}. This discrepancy originates from a small error in 't Hooft's original calculation \cite{tHooft:1976snw}, which was later corrected in an Erratum. However, the source for the instanton density \cite{Callan:1977gz} cited in \cite{Agrawal:2017ksf} still contains this error.
\item In \cite{Agrawal:2017ksf} the $\rho$ integration was cut off at $\rho = 1/M$ by hand. However, when working in the constrained instanton framework the $\rho$ integral is convergent and we find that the actual cutoff is roughly $\rho \sim 1/(2\pi v_\Sigma )$ (see also \cite{Fuentes-Martin:2019dxt}). 
\end{itemize}
The actual size of the suppression depends on the relation between the matching scale $M$ and the VEV $v_\Sigma$. Since $M$ corresponds to the mass scale of the massive gauge bosons, it scales like $M = g_{\rm eff}\,  v_\Sigma$, where $g_{\rm eff}$ is some combination of $g_1,\ldots , g_k$. For couplings of $\mathcal{O}(1)$ this leads to a suppression of $(2\pi)^{4-b_i}$, which is strongest for the $SU(3)$ group factors that do not couple to fermions. As we will show momentarily, this suppression is significant in the minimal model with only two group factors, but is less important once more $SU(3)$ factors are included and the matching relation in Eq. (\ref{eq:matching}) allows larger couplings in the individual $SU(3)$ factors.

\subsection{Example \texorpdfstring{$SU(3)^2, SU(3)^3\rightarrow SU(3)_{QCD}$}{SU(3)xSU(3),SU(3)xSU(3)xSU(3)-->SU(3)}}
\label{sect:instCompOrig}
Let us now have a look at the minimal model with $k=2$. In this case the matching scale is directly set by masses of the heavy gauge bosons (and scalars)
\begin{equation}
M^2 = (g_1^2+g_2^2) v_\Sigma^2\,.
\end{equation}
In order to do the matching we use the RG equation to run the $\overline{\text{MS}}$ QCD coupling from the top mass at $\alpha_s (m_t)=0.10$ to the matching scale $M$. The small instanton contribution to the axion mass relative to the QCD contribution can now be computed using the mass scales $m_{a_2}^2 f_{a_2}^2$ and $m_{a_1}^2 f_{a_1}^2$  from Eqs.~(\ref{eq:Lambdak}) and (\ref{eq:Lambda1}) respectively.
 For simplicity we will assume that $f_{a_1} = f_{a_2} =f_a$ and use Eq.~(\ref{eq:maxion}) to obtain a numerical value for $f_a m_a = (75.5\text{ MeV})^2$.
\begin{figure}
\centering
\subfigure[\label{subfig:axionMassMinimalModela}]{\includegraphics[width=0.47\textwidth]{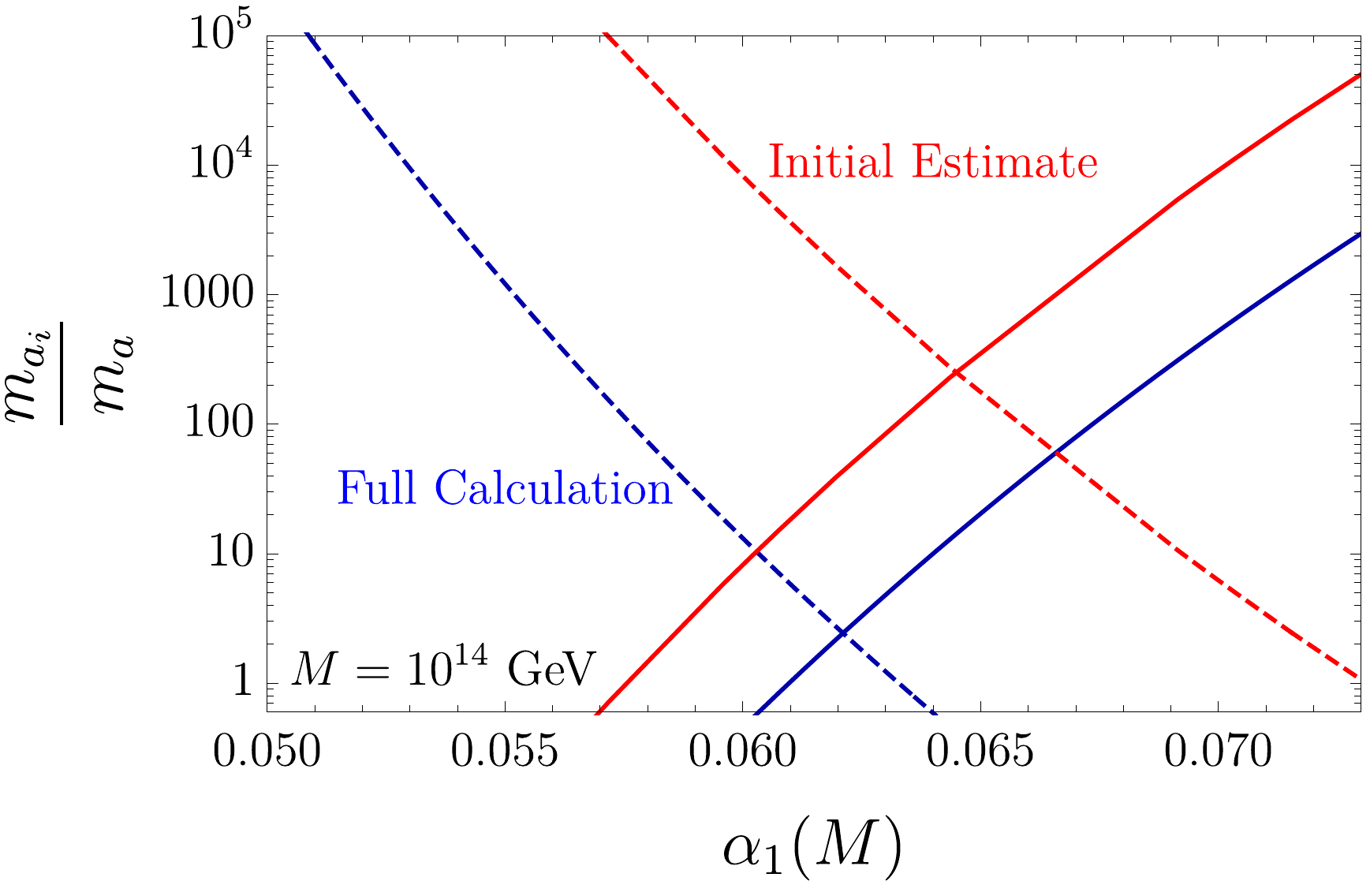}}\hfill
\subfigure[\label{subfig:axionMassMinimalModelb}]{\includegraphics[width=0.49\textwidth]{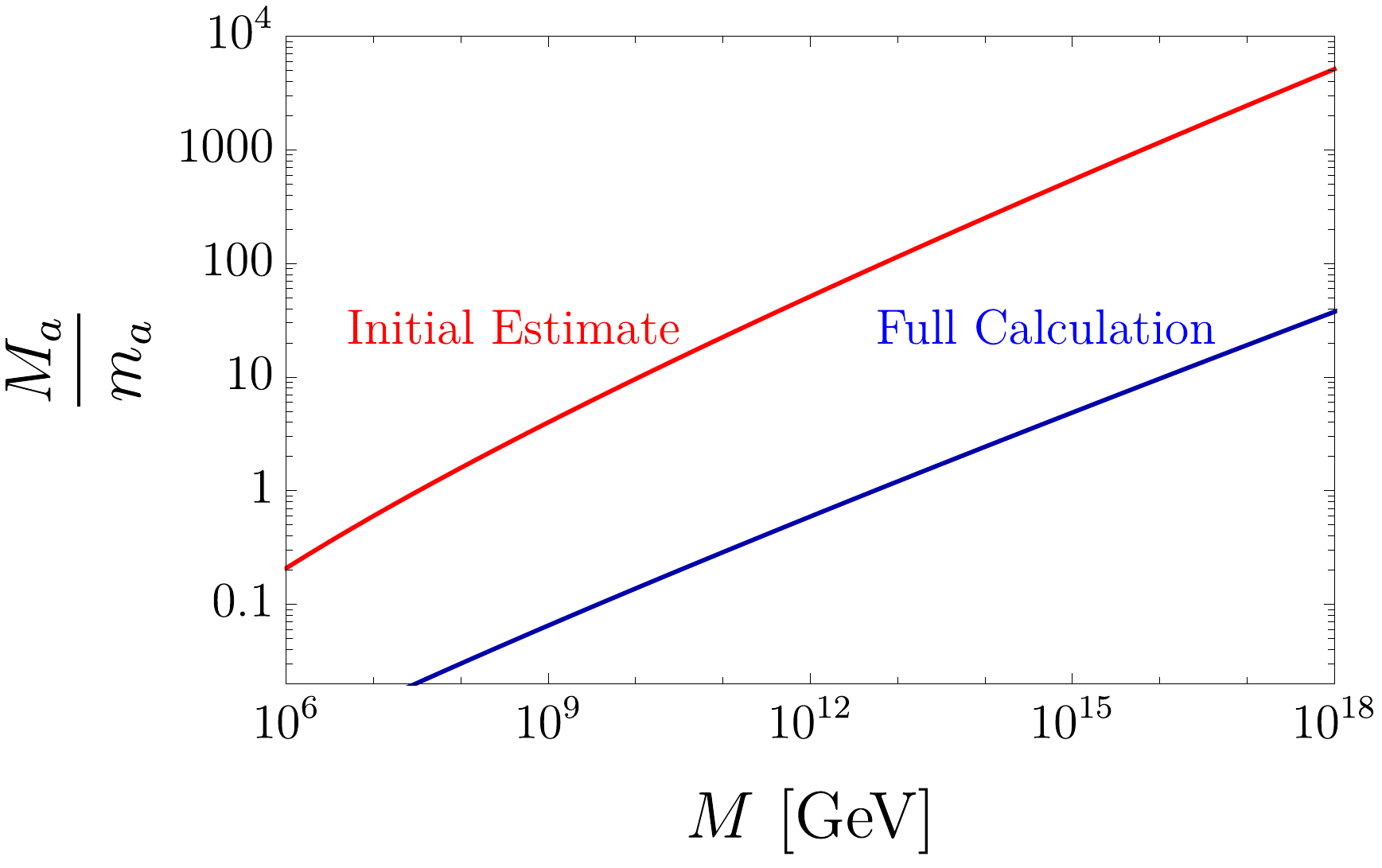}} 
\caption{Small instanton contribution to the axion mass relative to the IR QCD contributions in the model based on the symmetry breaking structure $SU(3)\times SU(3)\rightarrow SU(3)_{QCD}$. On the left we show our results for the full calculation in blue compared to the initial estimates in previous work \cite{Agrawal:2017ksf} in red at a breaking scale of $M=10^{14}$ GeV as a function of $\alpha_1 = \tfrac{g_1^2}{4\pi}$. The solid (dashed) curves show $m_{a_1}/m_a$ ($m_{a_2}/m_a$), which intersect at $M_a/m_a = 2.4$ in the full calculation and at $M_a/m_a = 251$ in previous estimates. On the right we show the values for $M_a/m_a$ at the intersection point of $m_{a_1}/m_a$ and $m_{a_2}/m_a$ for a wide range of breaking scales $M$. In both plots we took $f_{a_1} = f_{a_2} = f_a$.\label{fig:axionMassMinimalModel}}
\end{figure}
 
This ratio is shown for both axions (solid for $m_{a_1}/m_a$ and dashed for $m_{a_2}/m_a$) for the choice  of $M=10^{14}$ GeV for the symmetry breaking scale 
in Fig.~\ref{subfig:axionMassMinimalModela}.
 In contrast to previous estimates \cite{Agrawal:2017ksf} (shown in red), the full calculation shows that there is no region in parameter space where both axion masses are enhanced by more than an $\mathcal{O}(1)$ factor compared to the pure QCD prediction at the same time. One of the axions might be heavy, but then the other will be dominated by the QCD contribution to its potential and will therefore be like the standard QCD axion. The largest effect of small instanton contributions to both axion masses is found at the intersection of the two curves where both axions have the same mass which is about $M_a/m_a=2.4$ times heavier than the standard QCD axion. 
Fig. \ref{subfig:axionMassMinimalModelb}
 shows the maximal enhancement of the axion mass  due to small instantons as a function of the symmetry breaking scale $M$. This shows that even taking $M$ to be at the Planck scale the axion mass cannot deviate by more than a factor of $\sim 100$ from the QCD prediction. Due to the suppression factor in Eq. (\ref{eq:suppressionFactor}) the enhancement is  lower by about two orders of magnitude than the initial prediction in~\cite{Agrawal:2017ksf}.

We can therefore conclude that it is hard to get significant contributions from small instantons to the axion mass in the minimal model. However, according to our parametric estimate in Section \ref{sec:instantonEstimate}, we expect a larger mass enhancement in models with more $SU(3)$ factors. In the following we demonstrate that this conclusion is indeed correct by considering the next to minimal model with $k=3$ factors.

In the model with $k=3$ group factors $SU(3)^3$ is broken by the VEVs of two link fields, which we both take to be $\langle \Sigma\rangle = v_\Sigma\, \mathbb{1}_3$. Note that since $SU(3)_2$ couples to both link fields, not all gauge bosons get the same masses. One linear combination, corresponding to the QCD gluons, stays massless as before, whereas the masses of the other two linear combinations are given by
\begin{equation}
M_{V_{1/2}}^2 = \frac{v_\Sigma^2}{2} \big( g_1^2 + 2 g_2^2 +g_3^2 \pm \sqrt{4g_2^4 + (g_1^2-g_3^2)^2}\big)\,.
\end{equation}
The matching threshold is given by the geometric mean of these two mass eigenvalues 
\begin{equation}
 M=(g_1^2g_2^2+g_2^2g_3^2+g_1^2g_3^2)^{1/4}v_\Sigma\,.
\end{equation}

\begin{figure}
\centering
\subfigure[\label{subfig:axionMassCubedModela}]{\includegraphics[width=0.48\textwidth]{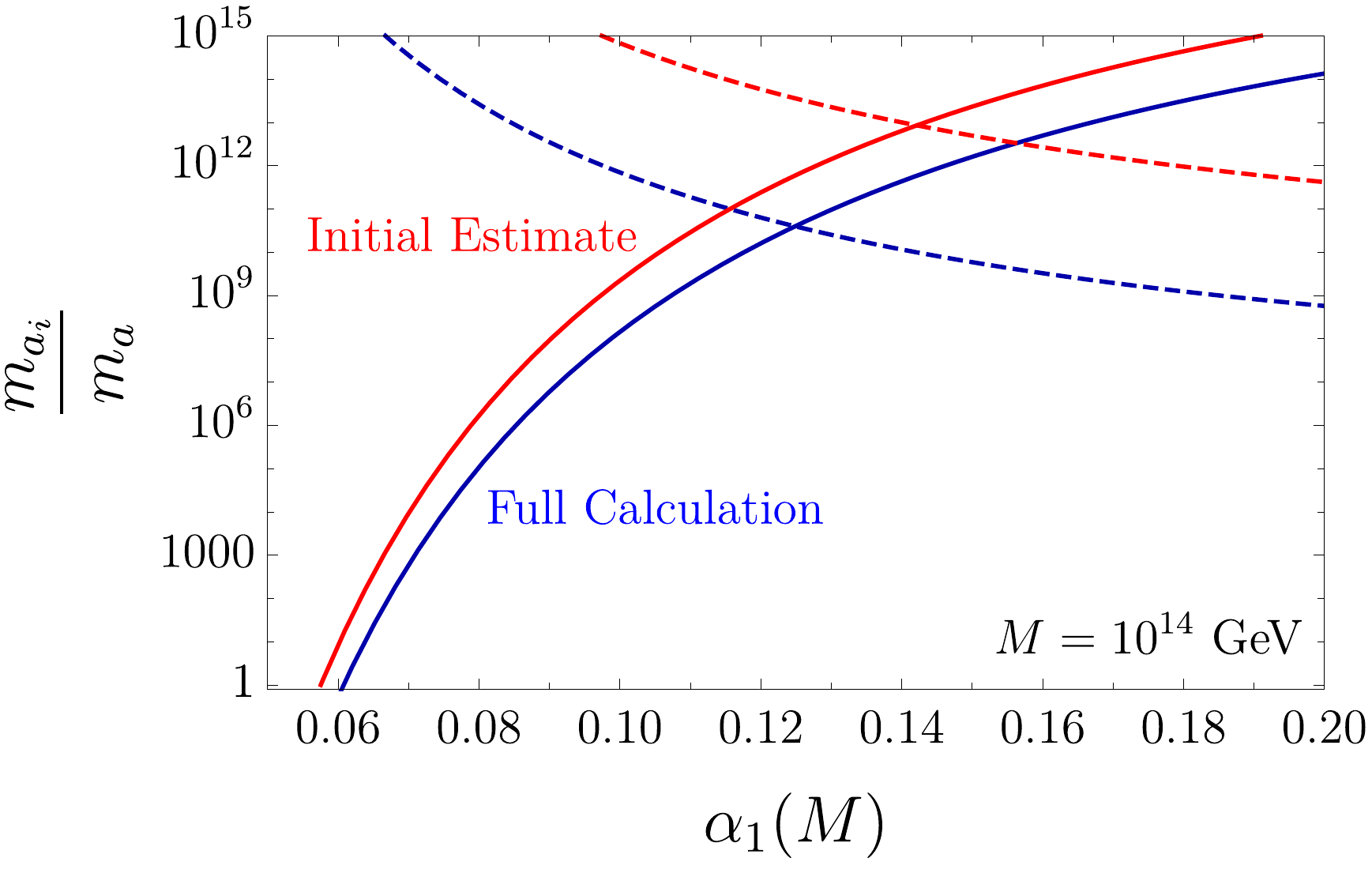}}\hfill
\subfigure[\label{subfig:axionMassCubedModelb}]{\includegraphics[width=0.488\textwidth]{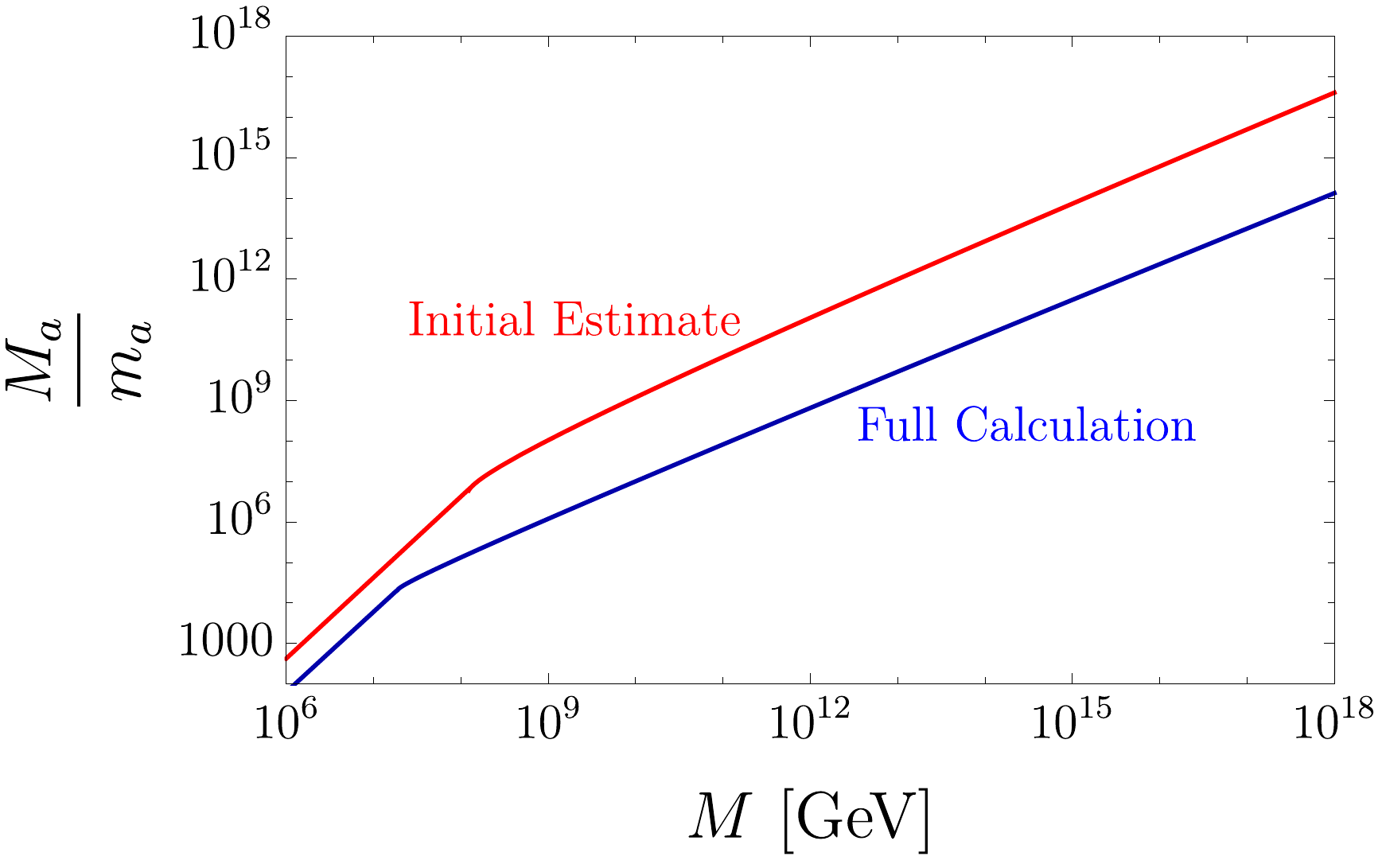}} 
\caption{Small instanton contribution to the axion mass relative to the IR QCD contributions in the model based on the symmetry breaking structure $SU(3)\times SU(3)\times SU(3)\rightarrow SU(3)_{QCD}$. On the left we show the results for the full calculation in blue compared to the initial estimates in previous work \cite{Agrawal:2017ksf} in red at a breaking scale of $M=10^{14}$ GeV as a function of $\alpha_1 = \tfrac{g_1^2}{4\pi}$. The solid (dashed) curves show $m_{a_1}/m_a$ ($m_{a_2}/m_a$), which intersect at $M_a/m_a \simeq 4\cdot 10^{10}$ in the full calculation and at $M_a/m_a = 9\cdot 10^{12}$ in previous estimates. On the right we show the values for $M_a/m_a$ at the intersection point of $m_{a_1}/m_a$ and $m_{a_2}/m_a$ or at the maximum of $m_{a_1}/m_a$ if they do not intersect for a wide range of breaking scales $M$. In both plots we took $f_{a_1} = f_{a_2} = f_{a_3} = f_a$ and fixed $g_2=g_3$, which implies that $m_{a_3}$ is always slightly larger than $m_{a_2}$. \label{fig:axionMassCubedModel}}
\end{figure}

As in the minimal model we take $f_{a_1}=f_{a_2}=f_{a_3}=f_a$ and show our result (blue) in Fig.~\ref{fig:axionMassCubedModel} for the small instanton contributions to the axion mass compared to  the estimates in~\cite{Agrawal:2017ksf} (red), fixing in both cases $g_2 = g_3$. 
In Fig.~\ref{subfig:axionMassCubedModela} 
we again show $m_{a_1}/m_a$ (solid) and $m_{a_2}/m_a$ (dashed) at a breaking scale of $M=10^{14}$ GeV. Note that $m_{a_3}$ is always larger than $m_{a_2}$ for identical couplings, since $m_{a_2}$ is suppressed by an additional factor of $2^{2-b_2/2}$. As can be seen, even though the mass enhancement is again smaller in the full calculation than in the initial estimate, small instantons can still enhance the mass of all three axions simultaneously by up to a factor of $4\cdot 10^{10}$ compared to the QCD contribution at the intersection point. 
Fig.~\ref{subfig:axionMassCubedModelb} 
shows that small instantons give dominant contributions to the axion mass also at smaller breaking scales $M$, making the axion considerably heavier than in the standard QCD axion scenario. Note that at small $M$, $m_{a_1}/m_a$ and $m_{a_2}/m_a$ do not intersect anymore. When this is the case $m_{a_1}/m_a < m_{a_2}/m_a $ due to its suppression by the Yukawa couplings and therefore we take the maximum of $m_{a_1}/m_a$ as an estimate for the maximal simultaneous enhancement of all axion masses. This is the reason for the kink in the curves in the  Figure \ref{subfig:axionMassCubedModelb}.

Adding additional $SU(3)$ factors increases the possible enhancement of the axion mass even further. It was already noted in \cite{Agrawal:2017ksf} that for $k \gg 1$ the axion masses scale as $m_{a_i} \sim M^2 / f_{a_i}$ for $i=2,\ldots ,k$ and $m_{a_1} \sim \sqrt{K} M^2 / f_{a_1}$, where the first axion mass is parametrically suppressed relative to the others by the Yukawa couplings and loop factors $\sqrt{K} \approx 10^{-12}$. With the help of Eq. (\ref{eq:smallinstwithk}) we can now understand the scaling of the axion mass with $M^2$ as the limit $b_{QCD}/ k \xrightarrow{k\rightarrow \infty} 0$ in Eq. (\ref{eq:smallinstwithk}).

\section{Conclusions\label{sec:Conclusions}}

We have presented a full calculation of the effects of small instantons on the axion mass in product group extensions of QCD. We found that a non-trivial embedding of the QCD instanton into the UV group will lead to an unusual scaling of small instanton contributions, which will appear as fractional instantons from the low-energy point of view. This opens up the possibility for small instantons in partially broken gauge groups to dominate over the large QCD instantons and significantly raise the axion mass away from the usual $m_a^2 - f_a$ relation. 

We have carried out the full 1-instanton calculation of the vacuum-to-vacuum transition amplitude of the broken product gauge group theories. As a first step we calculated the 1-instanton contribution to the vacuum bubble for a fully broken bosonic $SU(N)$ theory by performing the integral over the bosonic zero modes and non-zero modes. Since the gauge group is broken the effects of large instantons are exponentially cut off, and the calculation can be reliably carried out. While the inclusion of fermions and their interactions is usually handled using a 't Hooft operator approximation, we were able to evaluate the effects of the fermionic modes along with the Higgs loops needed to close up the zero modes directly without resorting to the 't Hooft operator approximation. This has the advantage that the result is manifestly finite and does not require the introduction of a regulator via a cutoff (whose exact definition in simple estimates usually leads to some uncertainty on the exact numerical value of the corrections). 

Using this method we were able to perform the calculation in the full theory with product groups broken to the diagonal and verify the scalings expected from our simple estimates. While the numerical value of the enhancement is not significant for the simplest 2 product group extension, already for 3 group factors we can obtain a large enhancement of the axion mass.

\section*{Acknowledgments}
We thank Paddy Fox, Tony Gherghetta, Valya Khoze, Alex Pomarol and Stefan Stelzl  for useful discussions.  C.C., M.R. and Y.S. thank the MIAPP at TU Munich for its hospitality while this project was initiated. C.C. also thanks the Aspen Center for Physics and the KITP at Santa Barbara for their hospitality while this work was in progress. M.R. thanks the Cornell particle theory group for its hospitality while this research was in progress. Y.S. thanks the Aspen Center for Physics for the hospitality while this work was in progress. The research of C.C. was supported in part by the NSF grant PHY-1719877 and in part by the US-Israeli BSF grant 2016153, as well as a Humboldt Foundation research prize. The work of M.R. was partially supported by the Collaborative Research Center SFB1258, by the DFG under Germany's Excellence Strategy - EXC-2094 - 390783311 and by the Studienstiftung des deutschen Volkes. Y.S. was supported in part by the NSF grant PHY-1915005.

\appendix 

\section{'t Hooft operator approach\label{app:tHooft}}
\label{app:tHooftOperator}

In this appendix we compute the small instanton contribution to the vacuum energy or axion potential in the presence of massless fermions using the 't Hooft operator approximation and compare it to the full calculation in Sec.~\ref{subsec:VacEnergy}.

In a gauge theory with $F$ massless fermion flavors in the fundamental representation of $SU(N)$ the pure vacuum-vacuum amplitude in the instanton background vanishes and the instanton configuration only contributes to correlation functions in which each fermion flavor and chirality appears at least once, i.e. for example $\langle 0| \prod_f (\bar{\psi}_f \psi_f)|0\rangle_{\Delta Q=1}\neq 0$. The effect of the instanton can thus be captured by the 't Hooft operator, which is an effective $2F$ fermion operator of the form (see e.g. \cite{tHooft:1976snw})
\begin{equation}
-\delta\mathcal{L}^{F} = e^{-i\theta} \int\frac{d\rho}{\rho^5} C_N (\rho) \rho^{3F} \big(\kappa_{N}^{(N_f)}\big)^{i_1\cdots i_{2 F}}\, \det_{f, f'}\,  (\bar{\psi}_{R\, f}(x_0) \psi_{L\, f'}(x_0) )_{i_1\cdots i_{2 F}} + h.c.\,,
\label{eq:tHooftOperator}
\end{equation}
where the determinant goes over flavor indices and the hermitian conjugate results from the anti-instanton configuration. $C_N(\rho)$ is defined in Eq. (\ref{eq:CNrho}) and $\big(\kappa_{N}^{(N_f)}\big)^{i_1\cdots i_{2 F}}$ is obtained by computing the $2F$ fermion correlation function in the instanton background and matching the result to the above effective operator. Note that the integration over the instanton location inside $SU(N)$, for which we assumed that $\sum_{i=1}^2\sum_{n=1}^{3} |\langle \tilde{\phi}_{in}\rangle |^2$ inside $C_N (\rho)$ is independent of the instanton position, projects out all invariant contractions of the fermion $SU(N)$ indices $i_1,\ldots , i_{F}$. For one fermion flavor the matching is straightforward (see \cite{Shifman:1979uw} for an example in $SU(2)$ and $SU(3)$) and gives
\begin{equation}
-\delta\mathcal{L}^{F=1} = e^{-i\theta}\int\frac{d\rho}{\rho^5} C_N (\rho) \rho^3 \kappa_{N}^{(1)} \bar{\psi}_{R\, 1}(x_0) \psi_{L\, 1}(x_0) + h.c.\,,
\end{equation}
where we used that $\big(\kappa_{N}^{(1)}\big)^{i_1i_2} = \kappa_{N}^{(1)} \delta^{i_1 i_2}$ (for example for $SU(3)$: $\kappa_{3}^{(1)}=\tfrac{4\pi^2}{3}$).

Since we want to close the `t Hooft operator with Higgs loops, we are only interested in flavor diagonal $SU(N)$ contractions of the form $(\bar{\psi}_R \psi_L)^{F}$. Therefore we will consider the effective Lagrangian
\begin{equation}
-\delta\mathcal{L}^{F} \simeq e^{-i\theta}\int\frac{d\rho}{\rho^5} C_N (\rho) (\rho^3 \kappa_{N}^{(1)})^{F} \prod_{f=1}^{F}\bar{\psi}_{R\, f}(x_0) \psi_{L\, f}(x_0) + h.c.\,.
\label{eq:NftHooftOperator}
\end{equation}
Note that due to Fierz relations among $SU(N)$ invariants, the prefactor $(\kappa_{N}^{(1)})^{F}$ is not exact, but will deviate from the full prefactor by an $\mathcal{O}(1)$ factor.

Such a 't Hooft operator contributes to the axion potential if one closes the fermion legs with loops. The leading contribution arises from closing the operator with Higgs loops via Yukawa couplings to the fermions as shown in 
Fig.~\ref{subfig:tHooftopb}. 
 This is the case since the diagram only includes marginal couplings and therefore scales as $M_{\rm cut}^{3F}$ where $M_{\rm cut}$ is the cutoff for the divergent loop integrals. 

Focusing on $SU(3)$ and identifying $\theta = \bar{\theta}- a_1 /f_{a_1}$, we can match the resulting operator to the effective Lagrangian in Eq. (\ref{eq:effLagrangian}) to obtain $m_{a_1}f_{a_1}$
\begin{equation}
m_{a_1}^2 f_{a_1}^2 = 2 K \int\frac{d\rho}{\rho^5} C_3 (\rho) (4\pi^2 M^3 \rho^3)^{F}\,,
\end{equation}
where $K$ contains the Yukawa couplings and loop factors
\begin{equation}
K = \prod_{f=1}^F \frac{y_f}{4\pi}\,.
\end{equation}
Note that we canceled a factor $N=3$ from the sum over colors in the loop for each fermion flavor with the $3^{-F}$ from $(\kappa_3^{(1)})^F$. Computing the $\rho$ integral one obtains
\begin{equation}
m_{a_1}^2 f_{a_1}^2 = K \left. d_3 (M)\right|_{S=3,F}(4\pi^2)^{F} \Gamma\bigg[\frac{3F+b_0^{(1)}-4}{2}\bigg] \bigg(\frac{M}{2\pi v_\Sigma}\bigg)^{b_0^{(1)}-4} \bigg(\frac{M_{\rm cut}}{2\pi v_\Sigma}\bigg)^{3 F} M^4\,,
\label{eq:Lambda1tHooft}
\end{equation}
where $M$ is the matching scale for the couplings.

Comparing this result to Eq. (\ref{eq:Lambda1}), which was obtained by including the SM Higgs and Yukawa couplings directly in the path integral evaluation of the vacuum-vacuum amplitude, one finds that with $M_{\rm cut}$, the cutoff of the loop integrals, an additional scale appears. However, the exact definition of $M_{cut}$ is ambiguous and always introduces an uncertainty. Since $M_{\rm cut}$ enters $m_{a_1}^2 f_{a_1}^2$ with a large power, even $\mathcal{O}(1)$ changes in the definition of $M_{\rm cut}$ can have a significant impact on $m_{a_1}^2 f_{a_1}^2$. This ambiguity is removed in the calculation in Section \ref{subsec:VacEnergy}, since the result is manifestly finite.

Both methods are equivalent and therefore we can use the result from Sec.~\ref{subsec:VacEnergy} to infer the appropriate definition of $M_{\rm cut}$ for this process. We find that both approaches yield the same result, up to an $\mathcal{O}(1)$ factor, if one identifies $M_{\rm cut} \simeq v_\Sigma$, in nice agreement with our intuitive expectations.

\section{Converting results to \texorpdfstring{$\overline{\text{MS}}$}{MSbar} scheme\label{app:MSbar}}
All results in Section \ref{sec:InstBrokenSUN} were derived in the Pauli-Villars regularization scheme. However, in perturbative calculations dimensional regularization and the MS or $\overline{\text{MS}}$ scheme are more common. In this appendix we briefly summarize how to convert the results to these schemes.
 
Already in~\cite{tHooft:1976snw} 't Hooft showed that in order to convert the results to dimensional regularization one has to do the substitutions
\begin{align}
\ln\mu_0 &\rightarrow \frac{1}{4-n} - \frac{1}{2} \gamma + \frac{1}{2} \ln 4\pi\quad &\text{(zero-modes)}\,,\\
\ln\mu_0 &\rightarrow \frac{1}{4-n} - \frac{1}{2} \gamma + \frac{1}{2} \ln 4\pi -\frac{1}{2}&\text{(kinetic terms)}\,, \label{eq:convKin}
\end{align}
where the first substitution has to be made for the $\mu_0$ originating from gauge and fermion zero-modes and the second for $\mu_0$ from kinetic terms, i.e. from the non-zero modes and scalar fields.\footnote{In~\cite{tHooft:1976snw} 't Hooft found $-\tfrac{5}{12}$ instead of the $-\tfrac{1}{2}$ in Eq. (\ref{eq:convKin}). This mistake was noted by Hasenfratz and Hasenfratz \cite{Hasenfratz:1981tw} and reconciled the disagreement with Shore, who did the instanton calculation using dimensional regularization \cite{Shore:1978eq}. 't Hooft corrected the $-\tfrac{5}{12}$ in Eq. (B12) of \cite{tHooft:1986ooh} to $-1$. However, this was later again corrected by Shifman \cite{Shifman:2012zz} to the $-\tfrac{1}{2}$ we use in Eq. (\ref{eq:convKin}).} This substitution only affects the running coupling in the exponential
\begin{align}
-\frac{8\pi^2}{g^2(1/\rho)} &= -\frac{8\pi^2}{g_B^2(\mu_0)} + \ln (\mu_0 \rho) \big[ (4N -F) + (\tfrac{1}{3}F -\tfrac{1}{3} N -\tfrac{1}{6} S )\big]\label{eq:PVcoupling}\\
&\rightarrow \frac{8\pi^2}{g_B^2(n)} + \bigg(\ln \rho +\frac{1}{4-n} +\frac{1}{2}(\ln 4\pi -\gamma )\bigg) b_0  -\frac{1}{2} \bigg(\tfrac{1}{3}F -\tfrac{1}{3} N -\tfrac{1}{6} S \bigg)\,,
\end{align}
where we separated in Eq. (\ref{eq:PVcoupling}) the contributions to $b_0$ originating from zero modes (first bracket) from the ones from non-zero modes (second bracket). The renormalized coupling now depends on the renormalization scheme. Here we will consider MS and $\overline{\text{MS}}$ scheme which are define by
\begin{align}
\frac{8\pi^2}{g_{\text{MS}}^2(1/\rho)} &= \frac{8\pi^2}{g_B^2(n)} + \bigg(\ln \rho +\frac{1}{4-n} \bigg) b_0\,,\\
\frac{8\pi^2}{g_{\overline{\text{MS}}}^2(1/\rho)} &= \frac{8\pi^2}{g_B^2(n)} + \bigg(\ln \rho +\frac{1}{4-n} +\frac{1}{2}(\ln 4\pi -\gamma )\bigg) b_0\,.
\end{align}
Note that in the above we have to identify \cite{tHooft:1976snw}
\begin{align}
&g_B(n)\rightarrow g_{\text{MS}}(\mu)\qquad\text{and}\qquad \ln \rho +\frac{1}{4-n}\rightarrow \ln (\rho\mu)\,,\\
&g_B(n)\rightarrow g_{\overline{\text{MS}}}(\mu)\qquad\text{and}\qquad \ln \rho +\frac{1}{4-n} +\frac{1}{2}(\ln 4\pi -\gamma )\rightarrow \ln (\rho\mu)\,,
\end{align}
where $\mu$ is the renormalization scale in dimensional regularization. Thus to convert our results to $\overline{\text{MS}}$ scheme we have to replace
 \begin{equation}
e^{-8\pi^2/g^2(1/\rho) -C_2 N} \rightarrow e^{-\tfrac{1}{12}(2 F -S)} e^{-8\pi^2/g_{\overline{\text{MS}}}^2(1/\rho) -C^{\overline{\text{MS}}}_2 N}\,,
\end{equation}
with $C^{\overline{\text{MS}}}_2$ given by
\begin{equation}
C^{\overline{\text{MS}}}_2 = C_2 -\frac{1}{6}\,.
\end{equation}
Using this the instanton density in $\overline{\text{MS}}$ scheme is given by
\begin{equation}
\left.d_N^{\overline{\text{MS}}} (\rho)\right|_{F,S} = e^{-\tfrac{1}{12}(2 F -S)+ \tfrac{1}{6}N}\left.d_N (\rho)\right|_{F,S}\,.
\end{equation}

\end{document}